\documentclass{aa}
\usepackage{graphicx} 
\usepackage{txfonts}
\usepackage{natbib}
\bibpunct{(}{)}{;}{a}{}{,}

\begin{document}
  \title {Photometric study of selected cataclysmic variables \thanks {The Appendix is only
      available in electronic form at http://www.edpsciences.org. Table~\ref{t_ltb}, as well as Table~\ref{t_log_all} 
      and Fig~\ref{f_sin_V1193Ori} -- Fig~\ref{f_sin_Scl} of the Appendix, are also available at the CDS via 
      anonymous ftp to cdsarc.u-strasbg.fr (130.79.128.5)
      or via http://cdsweb.u-strasbg.fr/cgi-bin/qcat?J/A+A/.}}
  \author{C. Papadaki\inst{1,2} \thanks {\email{Christina.Papadaki@oma.be (CP)}}
    \and H.M.J. Boffin\inst{3}
    \and C. Sterken\inst{2}
    \and V. Stanishev\inst{4}
    \and J. Cuypers\inst{1}
    \and P. Boumis\inst{5}
    \and S. Akras\inst{5,6}
    \and J.~Alikakos\inst{5,7}
    }
  \institute{Royal Observatory of Belgium, Avenue Circulaire 3, 1180 Brussels, Belgium 
    \and Vrije Universiteit Brussel, Pleinlaan 2, 1050 Brussels, Belgium 
    \and European Southern Observatory, Karl-Schwarzschild-Str. 2,
  85738 Garching, Germany
  \and Physics Department, Stockholm University, AlbaNova University Centre,
   106 91 Stockholm, Sweden
    \and Institute of Astronomy \& Astrophysics, National Observatory of Athens, I. Metaxa \& V. Pavlou, 
    P. Penteli, 15236 Athens, Greece
  \and University of Crete, Physics Department, 71003 Heraklion,
  Crete, Greece
  \and Astronomical Laboratory, Department of Physics, University of
  Patras, 26500 Rio--Patras, Greece}
  \date{Received date / Accepted date}
  
  \abstract  
  {}
  {We present time-resolved photometry of five relatively poorly-studied
  cataclysmic variables: \object {V1193 Ori}, \object {LQ Peg}, \object {LD 317}, 
  \object {V795 Her}, and \object {MCT 2347-3144}.}
  {The observations were made using four 1m-class telescopes
  for a total of more than 250\,h of observation and almost 16,000
  data points. For \object{LQ Peg} WHT spectroscopic data have been analysed as well.}
  {The light curves show a wide range of variability on
  different time scales from minutes to months. 
  We detect for the first time a brightness variation of 0.05\,mag in
  amplitude in \object {V1193 Ori} on the same timescale as the orbital period,
  which we interpret as the result of the irradiation of the
  secondary. A 20-min quasi-periodic oscillation is also detected. 
  The mean brightness of the system 
  has changed by 0.5\,mag
  on a three-month interval, while the flickering was halved.
  In \object {LQ Peg} a 0.05\,mag modulation was revealed with a period
  of about 3\,h. The flickering was much smaller, of the order of
  0.025\,mag. A possible quasi-periodic oscillation could exist 
  near 30\,min. For this object, the WHT 
  spectra are single-peaked and do not show any
  radial-velocity variations. The data of \object {LD 317} show a decrease in
  the mean magnitude of the system. No periodic signal was detected
  but this is certainly attributable to the very large flickering
  observed: between 0.07 and 0.1\,mag.
  For \object {V795 Her},
  the 2.8-hour modulation, thought to be a superhump
  arising from the precession of the disc, is present. 
  We show that this modulation is not stable in terms of
  periodicity, amplitude, and phase. Finally, for \object {MCT 2347-3144}, a
  clear modulation is seen in a first dataset obtained in October
  2002. This modulation is absent in August 2003, when the
  system was  brighter and showed much more flickering.}
  {}

  \maketitle
  %%%%%%%%%%%%%%%%%%%%%%%%%%%%%%%%%%%%%%%%%%%%%%%%%%%%%%%%%%%%%%%%%%%%%%%%%%%%%
  %%%%%%%%%%%%%%%%%%%%%%%%%%%%%%%%%%%%%%%%%%%%%%%%%%%%%%%%%%%%%%%%%%%%%%%%%%%%%
  %%%%%%%%%%%%%%%%%%%%%%%%%%%%%%%%%%%%%%%%%%%%%%%%%%%%%%%%%%%%%%%%%%%%%%%%%%%%%
   %%%%%%%%%%%%%%%%%%%%%%%%%%%%%%%%%%%%%%%%%%%%%%%%%%%%%%%%%%%%%%%%%%%%%%%%%%%%%
  %%%%%%%%%%%%%%%%%%%%%%%%%%%%%%%%%%%%%%%%%%%%%%%%%%%%%%%%%%%%%%%%%%%%%%%%%%%%%

  \section{Introduction}
  Cataclysmic variables (CVs) are close binary systems with an orbital
  period ($P_{\rm orb}$) in the range of a few hours. They consist of a low-mass star,
  that fills its Roche lobe and transfers mass to its white-dwarf (WD)
  companion. In non-magnetic systems, an accretion disc (AD) forms around
  the WD, which can go through different phases of
  activity. These systems have been classified according to their
  photometric and spectroscopic properties. In dwarf novae (DN), changes in
  the state of the AD are responsible for the outbursts
  that are observed, with most typical ones reaching 2--5\,mag. On the
  other hand, nova-like (NL) CVs exhibit such a high mass-transfer rate that the
  disc is permanently stuck in outburst. The VY Scl stars
  are also included in the NL category. They reveal occasional
  reductions in brightness of the order of a few magnitudes, due to
  temporary lowering of the mass transfer rate.
  
  CVs play a crucial role in the understanding of
  several physical phenomena and in particular in the study of the
  origin of viscosity in ADs, which are ubiquitous in
  astrophysics. Their short $P_{\rm orb}$ makes them an ideal
  target for observational studies.
  CVs present a range of variability from a few seconds to
  several years and as such are very interesting for photometric
  campaigns. These variations are indicative of many different physical
  processes that must be understood. These processes are discovered through a variety of
  behaviours visible in the light curves, e.g. flickering,
  quasi-periodic oscillations (QPOs), outbursts, humps, superhumps,
  eclipses, etc. Thus, when observing a CV, one is almost guaranteed to
  see the system in a different state than it was before, whose comparison
  should allow us to comprehend the system better.
  
  However, there are systems which attract more attention than
  others. Thus, in 1999, a programme that started at the Royal
  Observatory of Belgium by one of us (HMJB), aimed at 
  photometrically following poorly-known CVs, mostly of the
  NL type. The programme was based on the use of 1m-class
  telescopes. In this paper, we present some of the results of this
  campaign. Data are presented for five stars, \object {V1193 Ori}, \object {LQ Peg}, LD
  317, \object {V795 Her}, and \object {MCT 2347-3144} and have been collected
  over the period 2002-2005 in four observatories. 
  
  The paper is organised as follows: in Sect. 2, we present our
  observations and data reduction methodology. Sect. 3 gives the data
  collected for each star individually and in Sect. 4, we discuss the results. 
	
  \section{Observations and data reduction}
  We performed our photometric observations at four different observatories: 
  at the South African Astronomical
    Observatory (SAAO), at the Observatorium Hoher List
    in Germany (OHL), and at two observatories in Greece: Kryoneri (KR)
  and Skinakas (SK). 
    At the SAAO we used the 1-m telescope equipped with a
  back-illuminated 1024x1024, 24-micron pixel STE4 CCD camera with a
  field of view (FOV) of $5\farcm3 \times 5\farcm3$. 
    At the OHL, we used the 1-m Cassegrain-Nasmyth telescope with the 2048x2048,
  15-micron pixel HoLiCam CCD camera with a 2-sided read-out. We used
  the focal reducer and an effective FOV of  $14\farcm1 \times
  14\farcm1$.  As one side occasionally encountered problems
    during the reading, we moved our targets to the
    other side, therefore decreasing the FOV to
    $7\arcmin \times 7\arcmin$.
    At KR, the 1.2-m Cassegrain telescope
    was used with a 512x512, 24-micron pixel CCD camera and a $2\farcm5 \times 2\farcm5$ FOV.
    Finally, we performed observations with the 1.3-m Ritchey-Cretien
    telescope at SK, giving a FOV of $8\farcm5 \times
    8\farcm5$ and used the 1024x1024 pixel SITe CCD camera.
    Most of the runs were unfiltered, unless noted. 
     
    The CCD frames were processed for bias removal and flat-field correction.
    Small sets of biases were taken at regular time intervals during the night and
    were then combined into one median bias frame, which was then subtracted from
    all images. In case of non-negligible drops in the CCD temperature, new
    biases were obtained at that point and neighbouring images were processed
    with the new combined bias frame. For each night we obtained a median flat frame. In case
    of bad weather during both evening and morning twilight, the flat frame of
    the previous or next observing night was used. 
    
    Aperture photometry was made using the IRAF package 
    {\it apphot}. The circular aperture used for the integration of the flux
    and the computation of the instrumental magnitudes was equal to
    $2\times{FWHM}$. When the nights were photometric and stable, a mean $FWHM$ was
    calculated, resulting in a fixed nightly aperture. In case of
    unstable sky quality and therefore great changes of the $FWHM$ during the 
    night, it was divided into parts and each time interval was treated
    independently by calculating a different aperture in each case. 
    Differential photometry was then applied to the resulting
    magnitudes of the CV and the selected comparison stars. 
    The
    behaviour of the differences of the magnitudes between the 
    comparison stars throughout the campaign was checked. Only
    those comparisons with stable differences were selected.
    
    The same reduction technique and light curve generation has been
    applied to all CVs analysed in this
    paper. Whenever any offset takes place it will be stated in the
    corresponding section.

  %%%%%%%%-----------------------------------------------------%%%%%%%%%%%%%%%%
  %%%%%%%%---------------------\object {V1193 Ori}-----------------------%%%%%%%%%%%%%%%%
  %%%%%%%%-----------------------------------------------------%%%%%%%%%%%%%%%%
  \section {Selected targets and results}
  %%%%%%%%%%%%%%%%%%%%%%%%%%%%%%%%%%%%%%%%%%%%%%%%%%%%%%%%%%%%%%%%%%%%%%%%%%%%%
  %%%%%%%%%%%%%%%%%%%%%%%%%%%%%%%%%%%%%%%%%%%%%%%%%%%%%%%%%%%%%%%%%%%%%%%%%%%%%
  \subsection{\object {V1193 Ori}}

  \object {V1193 Ori} was accidentally discovered in 1985 as a very blue variable star
  by M. Hamuy \citep{ham,maz}. 
  \citet{fil} obtained spectra that revealed H$\alpha$ and H$\beta$ emission 
  superposed on broad absorption lines.
  The classification of \object {V1193 Ori} as a NL CV
  of the UX UMa subtype was made by \citet{bon}, who conducted photometric and spectroscopic observations. 
  \object {V1193 Ori} showed irregular flickering 
  with a peak-to-peak amplitude approaching 0.2\,mag, typical of
  NLs. No QPOs, $P_{\rm orb}$, or eclipses were found. 
  The only definite spectral feature
  was a broad and shallow H$\beta$ absorption line with a central emission peak, 
  observed in other NLs. 
  One 3.6\,h photometric run  was obtained by
  \citet{war}. The light curve showed flickering with a 
  total range of 0.25\,mag and no
  coherent modulations or eclipses were detected. 
  
  A spectroscopic $P_{\rm orb}$ of 0.165\,d ($3.96\rm h\pm43\rm s$)
  was reported by \citet{rin2}. The average of their spectra in 1989 showed H$\alpha$ in
  emission, while slight absorption wings flanked H$\beta$ and \ion{He}{i} (5876\,\AA),
  positively contributing to the classification of \object {V1193 Ori} as a UX UMa
  star. Other \ion{He}{i} lines were also weakly seen in emission. From
  H$\alpha$ emission line-profile variations in 1988, \citet{rin2} suggested
  that this is the result of irradiation of the secondary. The
  variations in 1989 revealed red wing spikes that occurred twice per
  orbit at $\phi=0.2$ and 0.8.
  
  Recently, \citet{ak} reported three periodicities from their analysis
  of unfiltered photometric CCD  observations during 24 nights between
  November 2002 and January 2003. Those are 0.1430\,d, 0.1362\,d, and
  possibly 2.98\,d, and they are interpreted as the $P_{\rm orb}$, the
  negative superhump period $P^-_{\rm sh}$, and the precession period
  of the disc $P_{\rm prec}$, respectively.
  They believe that the 0.1430\,d periodicity, being consistent with the
  1d$^{-1}$ alias of the proposed $P_{\rm orb}$ of 0.165\,d
  \citep{rin2,pap} and exceeding its power by a factor
  of 5, should be considered as a refined value of the $P_{\rm orb}$. 
  %%%%%%%%%%%%%%%%%%%%%%%%%%%%%%%%%%%%%%%%%%%%%%%%%%%%%%%%%%%%%%%%%%%%%%%%%%%%%
  %%%%%%%%%%%%%%%%%%%%%%%%%%%%%%%%%%%%%%%%%%%%%%%%%%%%%%%%%%%%%%%%%%%%%%%%%%%%%
	\smallskip
	
    Our dataset consists of photometric observations conducted at the SAAO (5 nights) and at the OHL
     (10 nights spreading over 3 periods). A preliminary analysis of the
    results was presented by \citet{pap}. All observations were unfiltered,
    while the exposure time varied between 30 and 60\,s according to the
    telescope used and the atmospheric
    conditions. Table~\ref{t_log_all} gives the observing
    log.
    For the purpose of differential photometry seven comparison stars
    were selected, as indicated in
    Fig.~\ref{f_fov_all}: S1 (\object{[HH95] V1193 Ori-5},
     $V$=13.513), S2 (\object{[HH95] V1193 Ori-9}, $V$=13.829), S3
     (\object{[HH95] V1193 Ori-7}, $V$=13.920), and S4 (\object{[HH95]
     V1193 Ori-11},
    $V$=15.110) from \citet{hen}, as well as S5
     (\object{U0825\_01407649}), S6 (\object{U0825\_01402900}), and
    S7 (\object{U0825\_01404164}) from the USNO (United States Naval
     Observatory) catalogue at ESO (European Southern Observatory). 
    
    Given the fact that no filters were used, we were not able
    to perform any transformations to a photometric
    system. Nevertheless, we have applied the following procedure
    that yields the most reliable magnitudes in our case. 
    From our comparisons we have chosen S1--S4 for which \citet{hen}
    give their $V$ magnitudes. Based on their differential
    magnitudes, we could derive the magnitude of
     \object {V1193 Ori}. Moreover, by checking the behaviour of the
    magnitude differences between two comparison stars at the SAAO
    and OHL, we applied the appropriate magnitude shift and were
    in this way able to compare all light curves.
   
    \begin {table*}[!ht]
      \caption {Long-term behaviour of three CVs.}
      \label{t_ltb}
      \centering
      \begin{tabular}{lll|lll|lll}
	\hline\hline
	UT date & $\mu_{CV}$ & $\mu_{shift}$ & UT date & $\mu_{CV}$ & $\mu_{shift}$ & UT date & $\mu_{CV}$ & $\mu_{shift}$\\
	\hline
	22oct02$^a$ & 13.83(6) & 0.298(5) &  18jan04$^a$ & 14.34(8)  & 0.406(36)   & 06nov03$^b$ & 14.70(10) & 0.329(10)\\
	23oct02$^a$ & 13.87(7) & 0.293(8) &  10oct04$^a$ & 14.27(7)  & 0.397(25)   & 08nov03$^b$ & 14.81(7)  & 0.341(14)\\
	24oct02$^a$ & 13.89(8) & 0.299(6) &  12oct04$^a$ & 14.12(5)  & 0.399(11)   & 12jan05$^b$ & 16.15(13) & 0.328(6) \\
	25oct02$^a$ & 13.89(6) & 0.299(4) &  10jan05$^a$ & 14.75(7)  & 0.410(12)   & 24oct02$^c$ & 16.77(11) & 1.912(11)\\
	26oct02$^a$ & 13.89(6) & 0.302(5) &  14jan05$^a$ & 14.75(6)  & 0.434(9)    & 25oct02$^c$ & 16.94(9)  & 1.915(8)\\
	04nov03$^a$ & 14.43(7) & 0.412(12)&  11sep03$^b$ & 13.90(5)  & 0.332(14)   & 26oct02$^c$ & 16.98(9)  & 1.909(8)\\
	05nov03$^a$ & 14.48(6) & 0.415(12)&  13sep03$^b$ & 14.01(8)  & 0.327(12)   & 27oct02$^c$ & 16.80(7)  & 1.877(11)\\
	06nov03$^a$ & 14.46(6) & 0.415(13)&  14sep03$^b$ & 13.90(9)  & 0.332(11)   & 13aug03$^c$ & 15.86(6)  & 1.896(8)\\
	08nov03$^a$ & 14.47(7) & 0.425(25)&  04nov03$^b$ & 14.63(11) & 0.330(13)   & 16aug03$^c$ & 16.00(8)  & 1.903(4)\\ 
	17jan04$^a$ & 14.36(3) & 0.399(17)&  05nov03$^b$ & 14.67(10) & 0.329(10)   & 17aug03$^c$ & 15.99(9)  & 1.895(5)\\
	\hline
      \end{tabular}
      {\footnotesize 
	    \newline 
	    Notes: $\mu_{CV}$ is the mean nightly magnitude of our object ; $\mu_{shift}$ is the nightly mean of the shift
	    between two comparison stars ; a,b, \& c indices correspond to observing runs of \object {V1193 Ori}, \object {LD 317}, 
	    and \object {MCT 2347-3144}, respectively.\hfill}
    \end {table*}

    Table~\ref{t_ltb}
    shows $\mu_{CV}$, the system's mean nightly magnitude and the variation of the difference
     between the mean magnitudes of two comparison stars, $\mu_{shift}$.
    Even though the
    difference between the standards' mean magnitudes varies up to 
    $\approx$0.1\,mag,
    this is negligible compared to the much larger variability of
    \object {V1193 Ori}. Not much attention should be paid 
    to the significance of
    the system's shift in 2004 and 2005, since 
    all individual runs belonging to these sequences are of small
    duration compared to the $P_{\rm orb}$, so the
    system's mean magnitude could be misleading. This is also why
    only the runs of 2002 and 2003 will be used for interpretation, as will be mentioned later.
    %%%%%%%%%%%%%%%%%%%%%%%%%%%%%%%%%%%%%%%%%%%%%%%%%%%%%%%%%%%%%%%%%%%%%%%%%%%%%
    %%%%%%%%%%%%%%%%%%%%%%%%%%%%%%%%%%%%%%%%%%%%%%%%%%%%%%%%%%%%%%%%%%%%%%%%%%%%%
    
    In our analysis we have divided the observing sequences
    into three
    different datasets, taking the different sites and 
    brightness states into account. The first one contains the light curves obtained
    at the SAAO, the second one those of November 2003 at the OHL,
    and the third one the remaining obtained at OHL.
    In Fig.~\ref{f_sin_V1193Ori} the individual light curves can be seen in detail, while
    in the upper left part of each light curve, the mean $\sigma$ of the
    comparison stars is indicated for correlation with the corresponding
    $\sigma$ of \object {V1193 Ori}.
    
\begin{figure}
      \resizebox{\hsize}{!}{\includegraphics[width=8cm]{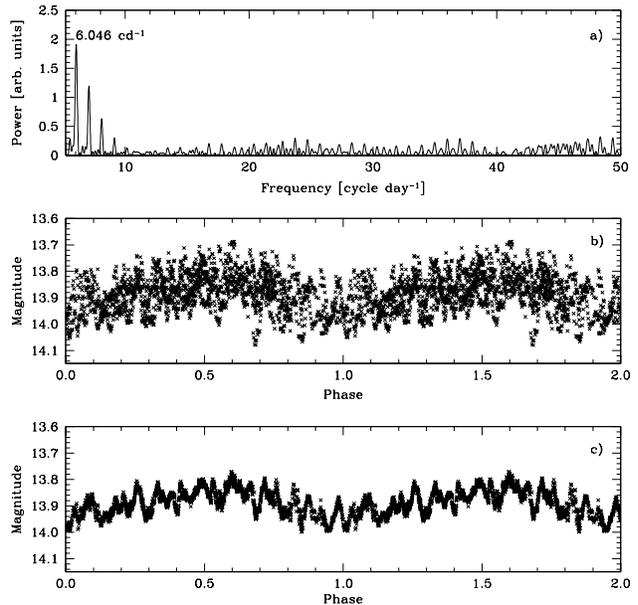}}
      \caption{a) Resulting PS of \object {V1193 Ori}, for
      the SAAO
      observing sequence, with the dominant frequency
      indicated. b) Data folded on the 3.97-h periodicity.
      c) Folded data, smoothed with a boxcar of 5.}
      \label{f_pof_V1193Ori}
    \end{figure}

    For this CV, as well as for the rest of this paper, frequency analysis
    was performed by using the period04 package \citep{len},
    based on the Discrete Fourier Transform method. Calculations of the
    uncertainties for the fitted parameters were performed by means of
    Monte Carlo simulations from the same package.
    For the first dataset, we confirm the
    previously reported spectroscopic $P_{\rm orb}$ of 3.96\,h. In more detail,
    we find a periodicity of 6.046(6)\,c\,d$^{-1}$ with a semi-amplitude of 0.047(2)\,mag.
    The corresponding frequency spectrum and the folded data on
    this periodicity can be seen in
    Fig.~\ref{f_pof_V1193Ori}a and Fig.~\ref{f_pof_V1193Ori}b,
    respectively. 
    To reduce the noise from random variability,
   the folded data have been smoothed in Fig.~\ref{f_pof_V1193Ori}c with a boxcar of 5.
    In the power spectrum (PS), the
    peaks surrounding the 6.046\,c\,d$^{-1}$ frequency
    are nothing but aliases, since they disappear as soon as this
    frequency is removed. Frequency analysis of the residuals did not reveal
    other coherent periodicities. The sinusoidal representation of the
    $P_{\rm orb}$ has been
    plotted in the light curves of Fig.~\ref{f_sin_V1193Ori}.
    %%%%%%%%%%%%%%%%%%%%%%%%%%%%%%%%%%%%%%%%%%%%%%%%%%%%%%%%%%%%%%%%%%%%%%%%%%%%%
    %%%%%%%%%%%%%%%%%%%%%%%%%%%%%%%%%%%%%%%%%%%%%%%%%%%%%%%%%%%%%%%%%%%%%%%%%%%%%
    
    For the second dataset, the same periodicity is still clearly
    visible only
    in the night of 4 November 2003. Its amplitude, though, has 
    decreased to 0.037\,mag. The rest of the sequences belonging to this
    dataset show no clear evidence of this periodicity. Its low amplitude, as
    well as the folding, leads us to the conclusion that it either
    does not exist any more or must have fallen below the detection limits.
    We note that none of the periodicities mentioned at \citet{ak} could be detected.
    Their application to the data gave amplitudes that were too small and unsatisfying foldings. 
     
    The third dataset consists of all the short
    runs obtained mainly in non-photometric conditions
    making it unsuitable for period analysis. However, the data seem to
    be consistent with the
    existence of a periodic signal. The sine-curve of the orbital frequency has been
    fitted individually to the last two datasets
    (Fig.~\ref{f_sin_V1193Ori}) only for comparison
    purposes and no conclusions should be drawn.
     
    In order to obtain the orbital ephemeris, the timings of the
    photometric maxima were determined by fitting a parabola to the
    corresponding peaks. We have only used the 6 maxima belonging to
    the runs in which the periodicity of 3.97\,h was detected. In this way
    we get:
    \begin {equation}
      T_{max}[HJD]=2452570.60069(352)+0^d.165001(4)E
    \end {equation}
     
    Had we not detected the periodicity in the first night of the second
    dataset, we would have believed that its detection only at the SAAO data
    could have been associated with how the system at that time
    was brighter than all the other datasets. In the first dataset it has
    a mean value of 13.9\,mag, in the second one 14.4\,mag, while in the
    third dataset it varies between 14.1 and 14.7\,mag. Even though the 
    periodicity exists in the data of 4 November 2003, its amplitude has decreased,
    so there could still be a possible correlation between the
    shifts in the system's brightness and the non-detection of the signal. 
    %%%%%%%%%%%%%%%%%%%%%%%%%%%%%%%%%%%%%%%%%%%%%%%%%%%%%%%%%%%%%%%%%%%%%%%%%%%%%
    \begin{figure}
      \resizebox{\hsize}{!}{\includegraphics[width=8cm]{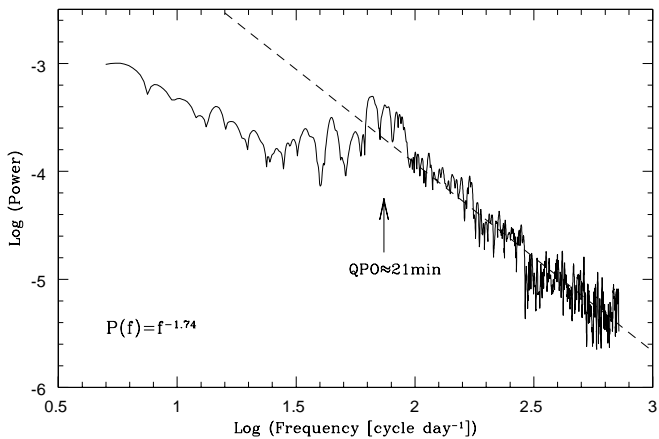}}
      \caption{The average PS of \object {V1193
	  Ori} in log-log scale. The dashed line
	  corresponds to the fit of
	  the linear part.}
      \label{f_log_V1193Ori}
    \end{figure}
    %%%%%%%%%%%%%%%%%%%%%%%%%%%%%%%%%%%%%%%%%%%%%%%%%%%%%%%%%%%%%%%%%%%%%%%%%%%%%
    
    We also searched for possible QPOs but no coherent signals
    were detected. Each night's PS showed
    many peaks, predominantly between
    50--100\,c\,d$^{-1}$. Since Fourier analysis could omit signals that are unstable in
    amplitude and frequency, the following procedure was applied. We averaged 
    the PSa of all nights, excluding the two shorter ones and plotted them in 
    log-log scale as shown in Fig.~\ref{f_log_V1193Ori}. A broad feature is 
    evident from 50\,c\,d$^{-1}$ on, with a peak around 70\,c\,d$^{-1}$. This feature
    is the counterpart of the QPO evident in our light curves. Inspection of the 
    individual average PSa of 2002 and 2003 revealed similar behaviour in the 
    same frequency range, albeit more prominent in 2003.   
    
    The PSa of CVs show so-called "red noise" characterised by 
    a power-law decrease of the power with the frequency (the power decreases linearly in log-log
    scale). It is usually  assumed that
    ``red noise'' is due to flickering through a shot noise-like
    process \citep[an extensive discussion of the flickering properties and origin in CVs can be
    found in][]{bru}. In the shot noise model, the light curve is
    generated by overlapping of many random
    "shots" (or pulses) of a given shape. If all shots have equal duration, the PS
    of the resulting light curve is simply the PS of the single pulses; i.e.
    it is determined by the shape of the individual pulses. For example, if the
    shots have an infinite short rise-time and then decay exponentially (sometimes called 
    classic "shot noise"), the power law index $\gamma$ (or the slope of the 
    linear part in log-log scale) is 2. If on the other hand the shots' durations are 
    different and follow some distribution, which is to be expected given 
    the suggested mechanisms 
    for generating  the flickering \citep{bru}, then $\gamma$ gets smaller. 
    Therefore, $\gamma$ can be changed in the shot noise model by at least two ways, 
    either by changing the distribution of the shots' duration or by 
    changing the shape of the shots \citep[see e.g.,][]{kra}. \citet{min}
     \citep[see also][]{tak} propose a model for the "red noise"
   from accreting compact objects, based on the self-organised criticality 
   concept of \citet{bak}, which could explain both the power-law shape of
    the PS and variability of the power-law index. We would like 
   to emphasise, however, that the  measurements of $\gamma$ 
    are very difficult to interpret and, despite the numerous measurements for many CVs, this 
    has so far not led to significant progress in our understanding of the physical origin of
    flickering. 
    Nevertheless, we measured $\gamma$ for the stars we study and specifically looked for 
    any seasonal variations that might indicate change of the flickering properties. 
    At least for one CV, \object{TT Ari}, \citet{kra} detected a decrease in $\gamma$ 
    and the activity of the flickering when the star switched from a "negative" to a "positive" 
    superhump state.
    
    It is clear from Fig.~\ref{f_log_V1193Ori} that "red noise'' is also present in the 
    mean \object {V1193 Ori} PS. The linear part from
    100\,c\,d$^{-1}$ on was fitted by a
    least-square linear fit to determine $\gamma = 1.74(1)$. From the average 
    log-log PSa of 2002 and 2003, $\gamma$ was found to be
    2.10(2) and 1.57(2), respectively. However, as discussed in \citet{kra}, these
    errors should be considered as a lower limit. The
    PSa were very noisy and small changes in the fitting
    interval caused changes in the value of $\gamma$ reaching 0.1 for
    the average PS and 0.15 for the individual ones. These errors
    should be considered as the real ones.
     
    The amount of flickering has also been
    measured by subtracting the sinusoidal
    periodicity from our light curves and measuring the $\sigma$
    in each light curve. Then the average of all $\sigma$
     and its $\sigma$ as uncertainty, were computed and found equal to 
     0.0584(120)\,mag.
     Moreover, from the light curves of Fig.~\ref{f_sin_V1193Ori}, we can see that the
     increase in the CV's mean magnitude by $\approx$0.5\,mag between the two
     first datasets has been followed by a decrease in the amplitude of the
     flickering activity. 
      The total range was found to be $\approx$0.2\,mag for the
     first dataset and 0.15\,mag or lower for the second, in agreement
     with previously reported values.   
    %%%%%%%%%%%%%%%%%%%%%%%%%%%%%%%%%%%%%%%%%%%%%%%%%%%%%%%%%%%%%%%%%%%%%%%%%%%%%   
    %%%%%%%%%%%%---------------------\object {LQ Peg}----------------%%%%%%%%%%%%
    %%%%%%%%%%%%-----------------------------------------------------%%%%%%%%%%%%
    %%%%%%%%%%%%%%%%%%%%%%%%%%%%%%%%%%%%%%%%%%%%%%%%%%%%%%%%%%%%%%%%%%%%%%%%%%%%%
    \subsection{\object {LQ Peg}} 
    \object {LQ Peg} (\object{PG 2133+115}) was first discovered in
    the Palomar-Green (PG) survey
    \citep{gre} and identified as a CV 
    in 1984 \citep{fer}. The same authors argue that \object {LQ Peg}
    is a thick-disc CV probably of the UX UMa subtype of NL CVs. 
    
     \citet{rin1} performed a radial-velocity study of \object {LQ Peg}
    that yielded a period of 2.9\,h, identified as the $P_{\rm orb}$.
    This result, though, does not stand on solid ground because of the weak emission
    lines that makes the velocity measurements difficult. Private
    communication with Dr. Ringwald clarified that there was indeed a
    very significant false-alarm probability for the $P_{\rm orb}$
    determination, a highly noisy periodogram, as well as insufficient spectra
    sampling. Until a new radial-velocity study has been done, we
    therefore must assume that the $P_{\rm orb}$  of \object {LQ Peg} is not known.  
    
    Interest in \object {LQ Peg} began
    to rise after the discovery of dramatic fadings of up to 5\,mag
    at irregular intervals. The first one was recorded on
    photographs taken in 1969 \citep{sok}, a second one in
    1999 \citep{wat, kat}, and recently a third one, albeit poorly
    sampled \citep{hon2}. Judging by the
    general pattern and the time-scale of the recovery from  minimum,
     \citet{kat} placed \object {LQ Peg} among the VY Scl subgroup
    of NL CVs.
    
    Time-resolved photometry of \object{LQ Peg} is scarce.
     \citet{mis} obtained a single-night 6-h photometry.
    The data showed no periodicity and there was fairly strong
    flickering. \citet{sch} conducted a photometric study
    during the star's return to normal brightness on eight nights in
    1999--2000 \citep{sch}. Considerable flickering was present
    during all nights (0.09 to 0.02\,mag) with a decline in amplitude as the system
    brightened. But again, no coherent periodicity was found.
   \smallskip
   
    Photometric observations of \object {LQ Peg} were conducted at SK and
    KR observatories. The exposure times
    varied from 45 to 60\,s depending on the telescope, the filter
    used (if any), and the atmospheric conditions. As indicated in Table~\ref{t_log_all}, we
    observed for 10 nights in total, from early June
    till late August 2004. For the SK observations, we used the
    Johnson-$R$ filter, while the KR ones were unfiltered.    
     %%%%%%%%%%%%%%%%%%%%%%%%%%%%%%%%%%%%%%%%%%%%%%%%%%%%%%%%%%%%%%%%%%%%%%%%%%%%%
    \begin{figure}
      \resizebox{\hsize}{!}{\includegraphics[width=8cm]{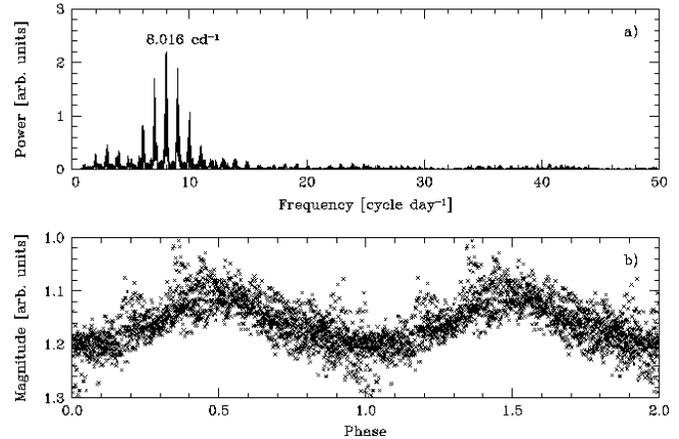}}
      \caption{a) \object {LQ Peg} PS, indicating the 2.99-h periodicity. b)
	The data of all the observing runs folded on this periodicity.}
      \label{f_pof_LQPeg}
    \end{figure}
    %%%%%%%%%%%%%%%%%%%%%%%%%%%%%%%%%%%%%%%%%%%%%%%%%%%%%%%%%%%%%%%%%%%%%%%%%%%%% 
     
    The comparison stars used to
    accomplish the differential photometry were always the same except
    at KR, where there were fewer due to the 
    smaller FOV of the camera, indicated with a box in Fig~\ref{f_fov_all}.
    The comparisons are: S1 (\object{[HH95]
    PG2133+115-2},
    $V$=13.544), S2 (\object{[HH95] PG2133+115-8}, $V$=13.097), S3
    (\object{[HH95] PG2133+115-23},
    $V$=15.132), S4 (\object{[HH95] PG2133+115-27}, $V$=13.916), S5
    (\object{[HH95] PG2133+115-31},
    $V$=14.888), S6 (\object{[HH95] PG2133+115-4}, $V$=14.693), S7
    (\object{[HH95] PG2133+115-35},
    $V$=14.692) from \citet{hen}, as well as S8 (\object{GSC 01128-00678}) from the
    General Sky Catalogue, and S9 (\object{U0975\_20472734}) and S10 (\object{U0975\_20473384}) from
    the USNO catalogue at ESO.
     A detailed view
    of the individual light curves with the night-to-night variations 
    removed and a constant 
    added to the arbitrary magnitude units  is shown in Fig.~\ref{f_sin_LQPeg}.
    In the upper left part of each light curve, the mean $\sigma$ of the
    comparison stars is indicated for comparison with the corresponding
    $\sigma$ of \object {LQ Peg}.
    %%%%%%%%%%%%%%%%%%%%%%%%%%%%%%%%%%%%%%%%%%%%%%%%%%%%%%%%%%%%%%%%%%%%%%%%%%%%%
    %%%%%%%%%%%%%%%%%%%%%%%%%%%%%%%%%%%%%%%%%%%%%%%%%%%%%%%%%%%%%%%%%%%%%%%%%%%%%
    
     Night-to-night variations in the system's mean magnitude may
    introduce false peaks in the periodograms. Prior to frequency
    analysis, we therefore removed those variations by subtracting
    its mean from each light curve. We detected a
    periodicity with frequency 8.0164(1)\,c\,d$^{-1}$ and semi-amplitude
    0.047(1)\,mag. This frequency corresponds to a periodicity of
    2.99\,h, and the resulting sine curve is shown in
    Fig.~\ref{f_sin_LQPeg}. 
    The PS, as well as the data folded on the detected periodicity, are given
    in Fig.~\ref{f_pof_LQPeg}a and Fig.~\ref{f_pof_LQPeg}b, respectively. 
    The 2.99-h periodicity does
    not only appear when the data are treated all together but even if the
    $R$-filtered and unfiltered runs are treated separately.   
     %%%%%%%%%%%%%%%%%%%%%%%%%%%%%%%%%%%%%%%%%%%%%%%%%%%%%%%%%%%%%%%%%%%%%%%%%%%%%
    \begin{figure}
      \resizebox{\hsize}{!}{\includegraphics[width=8cm]{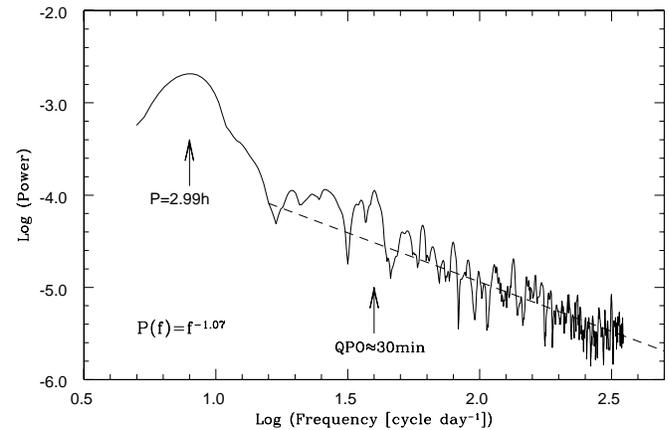}}
      \caption{\object{LQ Peg} average PS in log-log 
      scale. The dashed line corresponds to the fit of the linear part.}
      \label{f_log_LQPeg}
    \end{figure}
    %%%%%%%%%%%%%%%%%%%%%%%%%%%%%%%%%%%%%%%%%%%%%%%%%%%%%%%%%%%%%%%%%%%%%%%%%%%%% 

    Following the same procedure as for \object{V1193 Ori}, the amount
    of flickering and its corresponding $\sigma$ were found to be 0.0241(184)\,mag.
     The nights with the
    largest mean $\sigma$ of the comparison stars, were not
    used to derive the aforementioned value. The flickering source is also evident in
    Fig.~\ref{f_log_LQPeg}. It is characterised by
    $\gamma=1.07(3)$, while small changes in the fitting interval
    have given a real error of 0.15. The average PS of the
    June and August runs yielded  $\gamma=0.78(5)$ and
    $\gamma=1.35(5)$  with real errors, reaching 0.2--0.3. 
    
    A period analysis that was carried out on the
    residuals of the 2.99\,h revealed no coherent periodicities. 
    We have searched for possible QPOs by applying the same method as in Sec. 3.1. 
    In this case, a broad feature seems to exist between
    26--45\,c\,d$^{-1}$, as shown in Fig.~\ref{f_log_LQPeg}. However this 
    $\approx$30\,min candidate QPO has a power that only exceeds the continuum by a factor of
    $\approx$1.5. Therefore, its 
    reality cannot be secured.
     
    The observations cover 14 photometric maxima. After
     fitting a parabola to the corresponding
    peaks, the timings were determined and the following orbital ephemeris was obtained:
    \begin {equation}
      T_{max}[HJD]=2453158.55995(305)+0^d.124747(6)E
    \end {equation}
    
    We also extracted spectroscopic data
    from the Isaac Newton Group (ING) Archive. The spectra were
    obtained on the 15 August 1997 at the 4.2-m William
    Herschel Telescope (WHT) with ISIS, a high-efficiency, double-armed,
    medium-resolution spectrograph that permits simultaneous observing to
    be done in both red and blue channels. All the relevant information
    concerning the spectroscopic data can be found in Table~\ref{t_wht_LQPeg}.
    %%%%%%%%%%%%%%%%%%%%%%%%%%%%%%%%%%%%%%%%%%%%%%%%%%%%%%%%%%%%%%%%%%%%%%%%%%%%%
    \begin {table}
      \caption {Log of \object {LQ Peg} WHT Spectra.}
      \label{t_wht_LQPeg}
      \centering
      \begin{tabular}{lll}
	\hline\hline
	& BLUE & RED \\
	\hline
	UT Date & 15 Aug 1997 & 15 Aug 1997\\
	Grating & H2400B & R1200R\\
	Dispersion(\AA/pixel) & 0.11 & 0.41\\
	Total spectral range (\AA)& 442 & 420 \\
	$\lambda_{\rm central}$ (\AA) & 4798 & 6562 \\
	$HJD_{\rm start}$ & 675.576 & 675.591\\
	Duration(h) & 2.8 & 2.5\\
	\# Spectra & 14 & 10\\
	Exptime(sec) & 600 & 600 \\
	\hline
      \end{tabular}
    \end {table}
    %%%%%%%%%%%%%%%%%%%%%%%%%%%%%%%%%%%%%%%%%%%%%%%%%%%%%%%%%%%%%%%%%%%%%%%%%%%%%
    
    After bias subtraction, the spectra were processed, using the IRAF
    task {\it{doslit}}. Continuum division was then applied from the corresponding task
    {\it{continuum}}. No flat field correction or flux calibration was
    made. 
    We looked for radial-velocity variations in the emission lines
    but found variations only at the level below 10
    km\,s$^{-1}$ for H$\alpha$, i.e. below the spectral resolution in the
    red. We thus conclude that no radial-velocity variations
    are present within our limits.
    
\begin{figure}
      \resizebox{\hsize}{!}{\includegraphics[width=8cm]{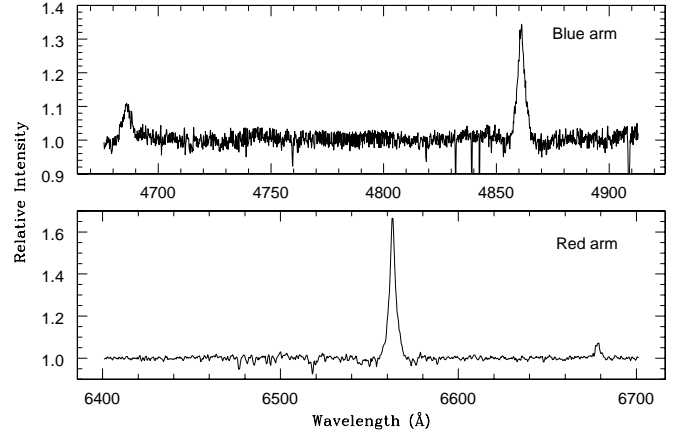}}
      \caption{Upper panel: \object{LQ Peg} median spectrum corresponding to the
	ISIS blue arm. Lower panel: median spectrum corresponding to the
	ISIS red arm.}
      \label{f_spe_LQPeg}
    \end{figure}

    The median
    combined spectra for the blue and red arm can be seen in the upper and lower
    panels of Fig.~\ref{f_spe_LQPeg}, respectively.
    %%%%%%%%%%%%%%%%%%%%%%%%%%%%%%%%%%%%%%%%%%%%%%%%%%%%%%%%%%%%%%%%%%%%%%%%%%%%%
    %%%%%%%%%%%%%%%%%%%%%%%%%%%%%%%%%%%%%%%%%%%%%%%%%%%%%%%%%%%%%%%%%%%%%%%%%%%%%
    %%%%%%%%%%%%%%%%%%%%%%%%%%%%%%%%%%%%%%%%%%%%%%%%%%%%%%%%%%%%%%%%%%%%%%%%%%%%% 
    They show emission lines identified as
    \ion{He}{ii} (4686\,\AA), H$\beta$, H$\alpha$, and \ion{He}{i} (6678\,\AA). As is evident
    from Fig.~\ref{f_spe_LQPeg}, all lines are weak compared to continuum, with only
    H$\alpha$ and H$\beta$ clearly visible in the individual spectra. No
    absorption lines were unambiguously detected. 
     The full widths and equivalent widths of the emission lines 
    are shown in Table~\ref{t_lin_LQPeg}; however, the
    values for \ion{He}{i} are quite uncertain since its flux was the lowest one,
    being close to the continuum with a profile that is not well-defined. 
    Only by computing the median spectrum did this peak reveal itself.
    %%%%%%%%%%%%%%%%%%%%%%%%%%%%%%%%%%%%%%%%%%%%%%%%%%%%%%%%%%%%%%%%%%%%%%%%%%%%%
    \begin {table}
      \caption {Characteristics of emission lines.}
      \label{t_lin_LQPeg}
      \centering
      \begin{tabular}{lll}
	\hline\hline
	& FW(km\,s$^{-1}$) & EQW(\AA) \\
	\hline
	\ion{He}{ii}(4686\,\AA) & 758 & 0.52 \\
	H$\beta$ & 784 & 1.46 \\
	H$\alpha$ & 889 & 3.65 \\
	\ion{He}{i}(6678\,\AA)& 550 & 0.24 \\
	\hline
      \end{tabular}
    \end {table}
    %%%%%%%%%%%%%%%%%%%%%%%%%%%%%%%%%%%%%%%%%%%%%%%%%%%%%%%%%%%%%%%%%%%%%%%%%%%%%   
     
    As mentioned before, \citet{fer} obtained a
    spectrum covering the range 3850--4950\,\AA, showing emission cores
    within the broad absorption lines of \ion{He}{ii} (4686\,\AA) and
    H. However, both \ion{He}{ii} (4686\,\AA) and H$\beta$ have appeared as
    pure emission lines in the WHT spectra we studied. The set of
    optical spectra of \object {LQ Peg} from \citet{rin1} also
    showed H$\alpha$ and \ion{He}{i} (6678\,\AA) in emission, but H$\alpha$ supposedly revealed
    radial velocities up to 300 km\,s$^{-1}$, in contrast to the
    present spectra. A closer inspection of
    a spectrum kindly sent to us by Dr. Ringwald revealed that the
    values 
    of the full widths of the two
    emission lines were almost double compared to ours. 
    %%%%%%%%%%%%%%%%%%%%%%%%%%%%%%%%%%%%%%%%%%%%%%%%%%%%%%%%%%%%%%%%%%%%%%%%%%%%%
    %%%%%%%%%%%%-----------------------------------------------------%%%%%%%%%%%%
    %%%%%%%%%%%%---------------------\object {LD 317}--------------------------%%%%%%%%%%%%
    %%%%%%%%%%%%-----------------------------------------------------%%%%%%%%%%%%
    \subsection{\object {LD 317}}
    \object {LD 317} was discovered by \citet{dah}
    based on approximately 68 magnitude estimates from
    photographic plates taken between 1967 and 1999. A spectrum obtained
    by Thorstensen confirmed it as a CV (www.cvcat.net), while H$\alpha$ spectroscopy
    (June 2001--October 2002) revealed a $P_{\rm orb}$ of 3.69\,h
    \citep{fen}. Although it is still uncertain
    uncertain, \object {LD 317} has been classified as a NL CV (www.cvcat.net).
    \smallskip
    
    Our photometric dataset consists of eight observing sequences in total,
    performed at the OHL. All runs were
    unfiltered, while the exposure time varied between 15 and 60 s,
    depending on the atmospheric conditions. A log of the details of all
    sequences can be seen in Table~\ref{t_log_all}.     
    
     The comparison stars used for the
    differential photometry are S1--S10, as shown in
     Fig.~\ref{f_fov_all}.
     Unfortunately, no well-defined magnitudes have been
     published for those present in our FOV. In
     order to obtain the transform equation, we used the $R$
     magnitudes of those comparisons just as they are given in the USNO
     catalogue at ESO. Those are S1 (\object{U1275\_18605964}, $R$=14), S2
     (\object{U1275\_18604470}, $R$=14.4), S3 (\object{U1275\_18605222}, $R$=14.4), S4
     (\object{U11275\_18600804}, $R$=14.4), S5 (\object{U1275\_18600027}, $R$=14.7), S6
     (\object{U1275\_18602067}, $R$=14.6), S7 (\object{U1275\_18602099}, $R$=14.4), S8
     (\object{U1275\_18604486}, $R$=15.3), S9 (\object{U1275\_18603877}, $R$=14.9), and
     S10 (\object{U1275\_18601201}, $R$=14.8).
    From each
    light curve we removed night-to-night variations by subtracting its mean
    We then added a constant to the arbitrary magnitude
    units and the result is shown in Fig.~\ref{f_lc_LD317}. In the upper left
    of each panel 
    in the same figure, the mean $\sigma$ of the
    comparisons' variations are presented through the error bar, which
    serves as a comparison to the system's variation.
    
    We performed a frequency analysis by 
    treating all light curves as a
    whole, as well as by dividing them into two sets. The first one consists of
    the observing runs of September 2003 and the second of November 2003.  
    This was done because the system between these two sets has faded by
    $\approx$1\,mag and a different behaviour could be expected. The reality
    of this fading can be seen in Table~\ref{t_ltb} when comparing $\mu_{CV}$ to the much smaller 
    shift between two comparison stars, $\mu_{shift}$.

   Nevertheless, no dominant
    periodicities were revealed. The only exception could be a periodicity of
    approximately 3.27\,h found at the two individual sets and one of
    3.48\,h
    found from all light curves together. These two could be related to
    the 2-d alias of the reported periodicity of 3.69\,h. However, in both
    cases, the amplitudes are low and the fit of the data to these
    periodicities is not very satisfying. 
    Our light curves (Fig.~\ref{f_lc_LD317}) indeed show
    random variability with the exception of the 13 and 14 September 2003
    sequences where a trend seemed to exist. 
    
    The flickering activity is found to differ between the
    observing runs of September and November 2003, where the system's mean
    magnitude has increased by almost 1\,mag. In September 2003 the
    flickering's mean value and corresponding $\sigma$ (computed in
    the same way as \object{V1193 Ori}) were found to be
    0.0689(51)\,mag. However, in November 2003 the flickering
    increased to 0.0949(177)\,mag.
    These light curves are dominated by much more random
    variability than before, as seen in Fig.~\ref{f_lc_LD317},
    making the detection of any modulation even
    more difficult. Furthermore, in January 2005 the system had
    further faded by 1.5\,mag with respect to
    November 2003. The
    light curve, however, still resembled that of November 2003, both in the
    amplitude of its flickering and in its random variability.
    To find the amount
    of flickering for November, the first short run was not used and
    the visible trend as seen in Fig.~\ref{f_lc_LD317} was
    removed from the remaining two. The rest of the runs were kept as they were. 
    Moreover, the power 
    law index of the flickering source's average PS was found to be  
    $\gamma=1.87(2)$. For September and November it was 1.44(2) and 1.86(2),
    respectively. Following the same procedure as before, the real
    error of the average PS approached 0.05 while of the individual
    ones  0.05--0.1. We also searched for possible QPOs but none was evident.
    %%%%%%%%-----------------------------------------------------%%%%%%%%%%%%%%%%
    %%%%%%%%---------------------V795  Her-----------------------%%%%%%%%%%%%%%%%
    %%%%%%%%-----------------------------------------------------%%%%%%%%%%%%%%%%
    %%%%%%%%%%%%%%%%%%%%%%%%%%%%%%%%%%%%%%%%%%%%%%%%%%%%%%%%%%%%%%%%%%%%%%%%%%%%% 
    \subsection{\object {V795 Her}}
     
    \object {V795 Her} (PG1711+336) was discovered in the PG survey \citep{gre}
    and identified as a CV candidate because of the
    weak hydrogen emission lines observed in its spectrum.
    Observations during the period 1983--2002 revealed a periodicity
    around 2.8\,h, occasionally vanishing below detection limits, as
    summarised in Table~\ref{t_2.8_V795Her}.
    At the beginning it seemed that each 
    value was just a more precise, ``corrected''
    one. But it appears that this modulation might
    not be such a stable ``clock'', its phase stability only being kept on
    a timescale of less than 20\,d \citep{pat1}. 
    Besides, \object {V795 Her} shows pronounced flickering activity 
    over all the years, as well as variations exhibiting a timescale of
    10--20\,min \citep{ros1}. QPO-like
    activity has also been reported at 15\,min \citep{ros2} and 1160\,s
    \citep{pat1}.
    %%%%%%%%%%%%%%%%%%%%%%%%%%%%%%%%%%%%%%%%%%%%%%%%%%%%%%%%%%%%%%%%%%%%%%%%%%%%%
    \begin {table}
      \caption {2.8\,h photometric modulation estimates for V795 Her.}
      \label{t_2.8_V795Her}
      \centering
      \begin{tabular}{cll}
	\hline\hline
	Years & Period (d) & Source\\
	\hline
	1983 & 0.115883 & \citet{mir}\\
	1983--1985 & 0.114488 & \citet{bai}\\
	1983--1985 & 0.1157550  & \citet{ros1}\\
	          & or 0.1158807 & \\
	1983--1988 & 0.1166728 & \citet{kal}\\
	1983--1989 & 0.1164865 & \citet{sha2}\\
	1983--1990 & 0.1164683  & \citet{zha}\\
	          & or 0.1164489 & \\
	1992--1994 & Absent & \citet{pat1}\\
	1992--1994 & Absent & \citet{cas}\\
	2002      & 0.116959 & \citet{pat2}\\
	\hline
      \end{tabular}
    \end {table}
    %%%%%%%%%%%%%%%%%%%%%%%%%%%%%%%%%%%%%%%%%%%%%%%%%%%%%%%%%%%%%%%%%%%%%%%%%%%%%
    
   Further periodicities have been detected for \object {V795 Her}
    (see Table~\ref{t_per_V795Her}). Among them lies
    a 2.6-h periodicity from the radial velocities of
    the emission lines \citep{sha2}, which is strongly
    believed to be the system's $P_{\rm orb}$.  
    %%%%%%%%%%%%%%%%%%%%%%%%%%%%%%%%%%%%%%%%%%%%%%%%%%%%%%%%%%%%%%%%%%%%%%%%%%%%%
    \begin {table}
      \caption {Additional periodicities reported for V795 Her.}
      \label{t_per_V795Her}
      \centering
      \begin{tabular}{cll}
	\hline\hline
	 & Type & Source\\
	\hline
	14.8\,h & Spectroscopic (Optical) & \citet{tho}\\
	2.60\,h & Spectroscopic (Optical) & \citet{sha2} \\
	     & Spectroscopic (Optical) & \citet{dic}\\
	     & Spectroscopic (HST UV)  & \citet{ros3}\\
	4.86\,h & UV resonance lines & \citet{pri1}\\
	     & UV resonance lines & \citet{pri2}\\
	
	\hline
      \end{tabular}
    \end {table}
    %%%%%%%%%%%%%%%%%%%%%%%%%%%%%%%%%%%%%%%%%%%%%%%%%%%%%%%%%%%%%%%%%%%%%%%%%%%%%
       
    \citet{tho} found an unusually long  spectroscopic
    periodicity of 14.8\,h, which he believed to be the $P_{\rm orb}$, but also no sign of the 2.8-h
    photometric signal. He proposed that, if both periods were true,
    then \object {V795 Her} could be a slow DQ Her.
    This opinion was partly abandoned when the 2.6-h
    spectroscopic period was detected \citep{sha1}. 
    \citet{sha2}, also in favour of an Intermediate Polar (IP)
    interpretation,
    came to the conclusion that the 2.6-h and not
    the 14.8-h is the $P_{\rm orb}$, while the 2.8-h photometric
    modulation is the beat period between the $P_{\rm orb}$ and the
    rotation period of the WD. However, the main disadvantages of the
    IP-model are that (1) X-ray observations have
    not revealed any of the periodicities reported, (2) the 
    expected rotational period of the WD has not been
    detected, and (3) the UV spectrum was found to be consistent with the properties of
    non-magnetic CVs rather than IPs \citep{pri1}. Nevertheless the IP
    model is not ruled out, taking into
    account a similar circumstance of the IP \object {EX Hya}.
    
    Another interpretation was put forward involving a precessing disc
    \citep{zha,pat1}.
    In this case, it is thought that 2.6\,h is $P_{\rm orb}$,
    while the 2.8-h modulation is the beat period
    between the $P_{\rm orb}$ and the $P_{\rm prec}$, corresponding in that way to a positive 
    superhump period $P^+_{\rm sh}$. The 2.8-h modulation is
    slightly unstable both in period and amplitude, just as
    superhumps are known to be. Last but
    not least, this interpretation is characterised by its physical
    plausibility. Theoretical studies suggest that
    ADs in binaries with a mass ratio $q<0.3$, which are large enough to
    make the tidal forces adequately strong, should indeed experience an
    eccentric instability. Considering the small $P_{\rm orb}$ of \object {V795 Her}
    that would lead to a small q \citep[around 0.2 as deduced in][]{warcv}
    and that discs are larger when
    the system is brighter and \object {V795 Her} is comparably bright to DN
    in outburst,together give firm ground for applying this theory \citep{pat1}. 
    
    \citet{cas}, who conducted medium-high resolution spectroscopy, 
    claim that \object {V795 Her} displays some of the key
    properties of SW Sex stars: Balmer line radial-velocity lag,
    attenuation of Balmer, and \ion{He}{i} flux around phase 0.5, accompanied by
    marked double peaks or central absorption. However, in contrast to
    the majority of SW Sex stars, it shows no eclipses and lies inside
    the period gap. A similar behaviour between
    the \object {V795 Her} emission lines and the ones of SW Sex
    stars was also noted by \citet{dic}.
    \smallskip
    
    Photometric CCD observations of \object {V795 Her} were performed
    at the KR, SK, and OHL observatories between
    June 2003 and May 2005. Depending on the telescope and the
    atmospheric conditions at that time, the exposure time varied between 20 and 100\,s. 
    We observed for 30 nights in total, of which 20 are
    unfiltered, 5 are in Johnson-$V$ filter, and 5 in Johnson-$R$ filter.
    Table~\ref{t_log_all} gives the observing log. 
     
    The comparison stars used to
    perform the differential photometry are indicated in
    Fig.~\ref{f_fov_all}, the box corresponding to the
    KR FOV. S1 (\object{[HH95] V795 Her-11}), S2
    (\object{[HH95] V795 Her-12}), S3
    (\object{[HH95] V795 Her-4}), S4 (\object{[HH95] V795 Her-5}) were
    used for SK and OHL, while S1--S2 for KR. 
     
    No transforms between
    the different runs were made, because
    different filters were used and a check of the difference
    between two comparison stars did not prove stable. Therefore, any
    comparison between light curves belonging to different 
    runs could be misleading. Consequently, further analysis will be
    performed after removing any night-to-night variations by
    subtracting from each
    light curve its mean. The resulting light curves, after adding a
    constant, are shown in Fig.~\ref{f_sin_V795Her}.
    %%%%%%%%%%%%%%%%%%%%%%%%%%%%%%%%%%%%%%%%%%%%%%%%%%%%%%%%%%%%%%%%%%%%%%%%%%%%%%%
    \begin{figure}
      \resizebox{\hsize}{!}{\includegraphics[width=8cm]{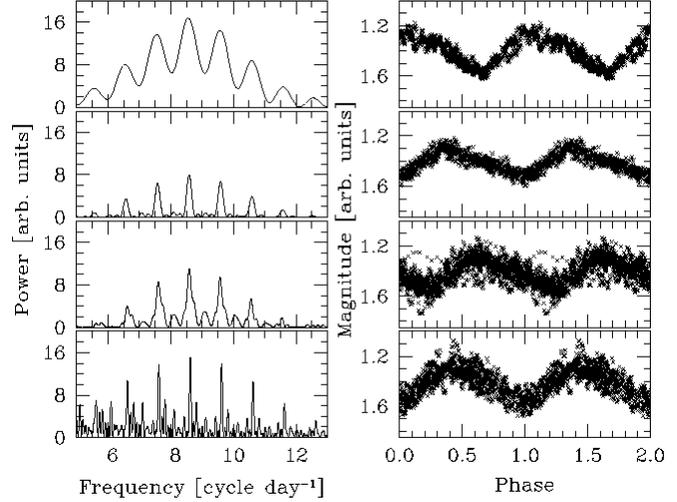}}
      \caption{From top to bottom: subgroups 1--4 of \object {V795 Her}, showing
	the PS on the left and the data folded on each subgroup's 
	periodicity (as shown in Table~\ref{t_sub_V795Her}) on the right.}
      \label{f_pof_V795Her}
    \end{figure}
    %%%%%%%%%%%%%%%%%%%%%%%%%%%%%%%%%%%%%%%%%%%%%%%%%%%%%%%%%%%%%%%%%%%%%%%%%%%%%
 %%%%%%%%%%%%%%%%%%%%%%%%%%%%%%%%%%%%%%%%%%%%%%%%%%%%%%%%%%%%%%%%%%%%%%%%%%%%%%%
    \begin{figure}
      \resizebox{\hsize}{!}{\includegraphics[width=8cm]{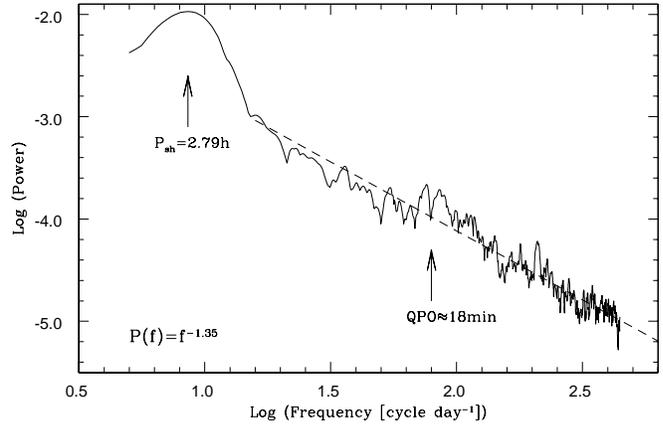}}
      \caption{\object {V795 Her} average PS in log-log 
      scale. The dashed line is the fit of the linear part.}
      \label{f_log_V795Her}
    \end{figure}
    %%%%%%%%%%%%%%%%%%%%%%%%%%%%%%%%%%%%%%%%%%%%%%%%%%%%%%%%%%%%%%%%%%%%%%%%%%%%%

    Using period04 \citep{len} for the whole dataset, we found a signal with 
     a frequency of 8.60238(1)\,c\,d$^{-1}$ corresponding to 2.79\,h or
    0.116\,d and a semi-amplitude of 0.083(1)\,mag.
     No eclipse was observed, while the
    pulse profile is asymmetric with a steeper rise than decline, as also
    noted by \citet{ros1}.

    We continued frequency analysis for the above residuals, and the
    next powerful frequency was 17.2\,c\,d$^{-1}$, the second harmonic of 8.602\,c\,d$^{-1}$,
    suggesting that the variation in our data is
    not completely sinusoidal, as is obvious from
   Fig.~\ref{f_sin_V795Her}. All subsequent frequencies did not
    have significant amplitude, and the corresponding folding revealed no trends.
        
    We divided our light curves into 4 subgroups of duration
    no longer than 20\,d and performed frequency analysis for each
    one. The two stand-alone runs of 5 June 2004 and 14 January 2005, which were
    shorter than 2.8\,h, have not been
    included. The same applies for the two short runs of September 2003,
    since the PS showed
    strong aliases and it was difficult to distinguish which peak was
    real. The
    subgroups can be seen in Table~\ref{t_sub_V795Her}. Included together are the
    corresponding highest peaks of the PS.
    %%%%%%%%%%%%%%%%%%%%%%%%%%%%%%%%%%%%%%%%%%%%%%%%%%%%%%%%%%%%%%%%%%%%%%%%%%%%%
    \begin {table}
      \caption {Frequency analysis of V795 Her individual subgroups.}
      \label{t_sub_V795Her}
      \centering
      \begin{tabular}{cccc}
	\hline\hline
	Subgroup & Time interval & $f$[c\,d$^{-1}$] & Amp.[mag]\\
	\hline
	1 & 02--03jun03 & 8.548(6) & 0.133(2)\\
	2 & 19--23jul03 & 8.586(2) & 0.090(2)\\
	3 & 14--21may04 & 8.589(2) & 0.104(2)\\
	4 & 03--14may05 & 8.628(2) & 0.125(3)\\
	\hline
      \end{tabular}
    \end {table}
    %%%%%%%%%%%%%%%%%%%%%%%%%%%%%%%%%%%%%%%%%%%%%%%%%%%%%%%%%%%%%%%%%%%%%%%%%%%%%
    It appears that the
    modulation appearing in all subgroups is indeed unstable
    both in amplitude and period. In this respect, a single period
    fitted to one subgroup is often out-of-phase with the best-fit
    sinusoid of the other subgroups. 
    This is clearly shown in 
    Fig.~\ref{f_pof_V795Her}, where each row from top to bottom represents subgroups 1--4.
    The PS from 5--13\,c\,d$^{-1}$ and the data folded on the corresponding periodicity are
    shown in the left and right columns, respectively. It was also
    noticed that, if taken at face value, the change in the
    period corresponds to a decrease of 0.005\,d over a year. Fig.~\ref{f_sin_V795Her} shows 
    the sinusoidal representation of each subgroup's periodicity  superimposed on the light curves  
    belonging to the same subgroup. The two individual nights of 5 June 2004 and 14 June 2005, as well as those
    of September 2003, were fitted with 
    the sine curves of the subgroups closer to them. 
    
    QPOs have been reported on in the past. Lately, \citet{pat2} reported a broad signal 
    near 80\,c\,d$^{-1}$ that possibly resolved
    into components at 71.0 and 80.4\,c\,d$^{-1}$ in 2002 and similar results but at 
    slightly different frequencies in 1993. We constructed the average PS
    of all runs exceeding a duration of 2\,h and confirm the existence of a 
    QPO near 80\,c\,d$^{-1}$ as is clearly shown in 
    Fig.~\ref{f_log_V795Her}. In our case the broad signal also resolves into 
    two components with slightly shifted frequencies compared to the ones reported by 
    \citet{pat2}. An inspection of each year's average PS revealed the 
    same QPO and its components, though each time at slightly different frequencies and 
    power. 
    
    The power law index that characterises the ``red noise'' seen 
    in  Fig.~\ref{f_log_V795Her} has a value of $\gamma=1.35(2)$. Its values for 2003, 2004, 
    and 2005 individually are 1.55(3), 0.84(5), and 1.33(4),
    respectively. The real errors coming from changes in
    the fitting region reach 0.17 for the average PS and
    vary between 0.25--0.4 for the average PSa of the individual years.
    The amount of flickering was also measured by
    removing the appropriate sinusoidal periodicity from each subgroup
    and following the same procedure as in the previous CVs. For all runs it was
    found to be 0.0526(124)\,mag.
    Additionally, the mean values
    for the 2004 and 2005 runs increased by 36\% in comparison
    to the 2003 runs, after treating each year individually,.  
 
    %%%%%%%%%%%%%%%%%%%%%%%%%%%%%%%%%%%%%%%%%%%%%%%%%%%%%%%%%%%%%%%%%%%%%%%%%%%%%
    %%%%%%%%-----------------------------------------------------%%%%%%%%%%%%%%%%
    %%%%%%%%---------------------\object {MCT 2347-3144}-------------------%%%%%%%%%%%%%%%%
    %%%%%%%%-----------------------------------------------------%%%%%%%%%%%%%%%%
    %%%%%%%%%%%%%%%%%%%%%%%%%%%%%%%%%%%%%%%%%%%%%%%%%%%%%%%%%%%%%%%%%%%%%%%%%%%%%
    \subsection{\object {MCT 2347-3144}}
    \object {MCT 2347-3144} was recently discovered and classified as a
    CV in the
    Montreal-Cambridge-Tololo (MCT) survey of southern sub-luminous blue
    stars \citep{lam}. This discovery did not undergo any observational
    follow-up or subsequent analysis. It has therefore been one of
    our selected targets of poorly known CVs.
    \smallskip
    
    Our dataset consists of two observing runs (totalling 7 nights) conducted at the
    SAAO. The first one was
    acquired in October 2002 and the other one in August
    2003. All runs were unfiltered, and the exposure time varied
    between 20 and 120\,s depending on the atmospheric conditions. The
    observing journal is given in Table~\ref{t_log_all}. A preliminary
    analysis of the results was presented by \citet{pap}.
      
    The two comparison stars indicated in Fig.~\ref{f_fov_all}, S1 (\object{U0525\_44380565}, $R$=14.7) and
    S2 (\object{U0525\_44381883}, $R$=16.6),
     were selected for differential
    photometry. Their magnitudes and designations
    were adopted from the USNO at ESO catalogue. The reality of the
    increase in the system's mean magnitude by $\approx$1\,mag in a one-year period
     is evident in Table~\ref{t_ltb} when comparing $\mu_{CV}$ to $\mu_{shift}$.
    
\begin{figure}
      \resizebox{\hsize}{!}{\includegraphics[width=8cm]{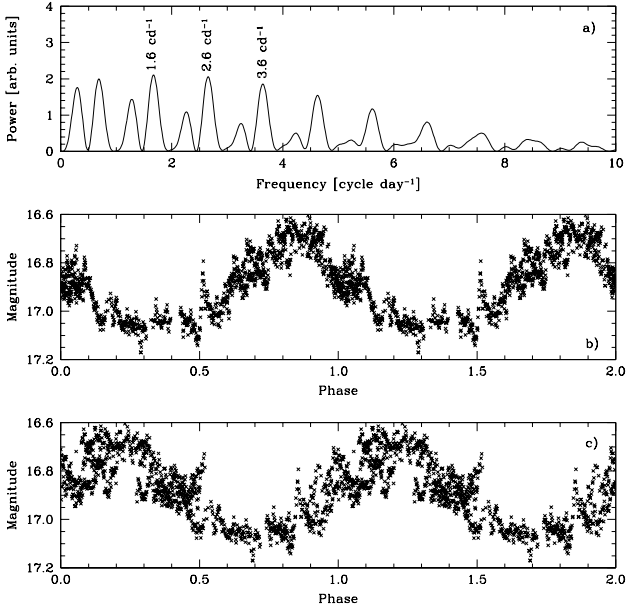}}
      \caption{(a) A close-up of the \object{MCT 2347-3144} power
      spectrum for the October 2002 runs, indicating the
      similarity of the highest peaks' power. (b,c) Folding of the
      same runs on 2.6\,c\,d$^{-1}$ and 3.6\,c\,d$^{-1}$, respectively.}
      \label{f_pfo_Scl}
    \end{figure}

    Frequency analysis was first performed on the October 2002 data. The most powerful
    frequency that was detected is 1.608(2)\,c\,d$^{-1}$ of
    semi-amplitude 0.182(3)\,mag along with its
    aliases. However, it was difficult to distinguish
    which should be the accepted frequency. Indeed, as
    seen in Fig.~\ref{f_pfo_Scl}a, the frequencies of 1.6\,c\,d$^{-1}$, as well as 0.6, 2.6, and
    3.6\,c\,d$^{-1}$, have very similar power. We therefore tested each one
    with the aim of choosing the one that best fits the data and has
    the smallest residuals as well. In this way we concluded
    on two most probable frequencies. Either 2.6\,c\,d$^{-1}$
    corresponding to 9.2 h
    or 3.6\,c\,d$^{-1}$ corresponding to 6\,h. The folding of the October 2002
    data on both periodicities can be found in Fig.~\ref{f_pfo_Scl}b
    and Fig.~\ref{f_pfo_Scl}c.
    The residuals of the two probable periodicities were
    searched further but no other periodicities were revealed.
    
    In contrast to the first dataset, the second one
    did not reveal any coherent periodicities. The light curves can be
    seen in more detail in Fig.~\ref{f_sin_Scl}. Since the mean
    $\sigma$ of the comparison stars, represented by error bars in the
    same figure, does not change for the two datasets, the large
    scatter visible in the second one should not be attributed to
    poorer weather conditions but to an increase of $\approx$50\% in the
    amplitude of the flickering.
    In particular, for 2003
    its mean value and $\sigma$ were 0.0771(39)\,mag, while for 2002 they
    were 0.0528(34)\,mag (computations based on the same
    procedure as for \object {V1193 Ori}), after
    subtracting the sinusoidal periodicity of 2.6\,c\,d$^{-1}$. Therefore, even if the
    modulation existed, we would not have been able to detect it.
    
    Because of the uncertainty in determining the exact
    periodicity in the first dataset, no ephemeris will be
    given. However, by fitting a parabola, we obtained the only
    time of maximum available, which was found to be 2452572.356\,d.
    %%%%%%%%%%%%%%%%%%%%%%%%%%%%%%%%%%%%%%%%%%%%%%%%%%%%%%%%%%%%%%%%%%%%%%%%%%%%  
    %%%%%%%%%%%%%%%%%%%%%%%%%%%%%%%%%%%%%%%%%%%%%%%%%%%%%%%%%%%%%%%%%%%%%%%%%%%%  
     \begin{figure}
      \resizebox{\hsize}{!}{\includegraphics[width=8cm]{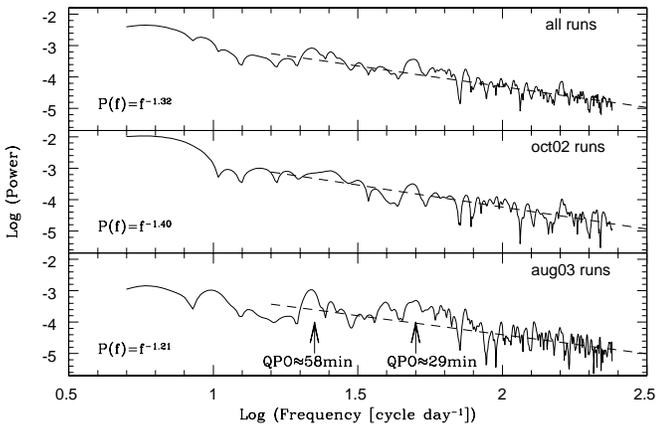}}
      \caption{From top to bottom: \object{MCT 2347-3144} average PSa in log-log
	scale for all runs, October 2002 runs, and August 2003 runs. The fit of the linear 
      parts of the PSa along with the corresponding equations are also shown.}
      \label{f_log_Scl}
    \end{figure}
    %%%%%%%%%%%%%%%%%%%%%%%%%%%%%%%%%%%%%%%%%%%%%%%%%%%%%%%%%%%%%%%%%%%%%%%%%%%%%
    
    In Fig.~\ref{f_log_Scl} the log-log PSa of all observing runs, as well as 
    of October 2002 and August 2003 alone, are shown. A broad signal
    near 25\,c\,d$^{-1}$ or 58\,min is clear in the 2003 graph. This broad signal appears to 
    resolve into two components, but is not seen in the runs of 2002. Moreover, 
    another broad signal could exist near 50\,c\,d$^{-1}$, but it is not as clear. Given the
    fact that there were not many runs to average and the non-persistence of the 
    signals during the 1-year period, we can only propose these two signals and especially
    the first more prominent one as candidate QPOs.  The power law
    index for the average PS, as well as the two sets of runs,
    was found 1.32(5), 1.40(8), and 1.21(8), respectively.
    Small changes in the fitting region of the linear part (in
    log-log) resulted in an error reaching 0.25 for the average PS and
    0.25--0.5 for the individual PSa of each year.
    
 \section{Discussion}
 \subsection{\object {V1193 Ori}}
   For the first time we confirm \object {V1193 Ori}'s $P_{\rm orb}$
   of 3.97\,h  photometrically. 
    The failure of the previous photometric campaigns to detect
    it can be attributed to a combination of
    reasons: the rapid high-amplitude flickering, the much smaller amplitude of the
    periodicity than the flickering, as well as
    the insufficient duration of the runs.
    We also note that no eclipses were seen in our light curves, in
    accordance to \citet{war} and \citet{bon}. 
    
    We find evidence of a QPO around 20\,min and confirm the high-amplitude 
    flickering. The power-law index $\gamma$ was found to be 1.74(1) for
    the average PS, while it differed between
    the individual average PSa of 2002 and 2003. It is interesting to
    note that in 2003 the decrease in the mean system brightness was
   followed by a decrease in the flickering amplitude and $\gamma$, 
    similar to \object{TT Ari}.
    As discussed in \citet{kra}, this might reflect either changes in the
    pulse shapes or a change in the distribution of the shots'
    duration. The same interpretation applies to 
    the rest of this work's CVs that have shown seasonal changes in
   $\gamma$. 
    It should also be noted that the decrease in $\gamma$
    between 2002 and 2003 is accompanied by the disappearance of the 
    periodic photometric modulation in 2003.
    
    A possible origin of the periodic brightness change is, given the morphology
    of the folded data, the low amplitude of the periodicity, and the
    fact that the observed signal 
    has a periodicity identical to the $P_{\rm orb}$, extra light of the side of the secondary that is face-on with
    the primary, i.e. the irradiation of the companion star. 
    One asset of this explanation is that Ringwald et al. (1994), who have
    performed the only spectroscopic study of \object {V1193 Ori} so far, also
    suggest that the H$\alpha$ emission line profile variations throughout
    the orbit could result from the irradiation of the secondary.  
    %%%%%%%%%%%%-----------------------------------------------------%%%%%%%%%%%%
	\subsection{\object {LQ Peg}}
    We have been able to detect a clear modulation of
    2.99\,h. Flickering was found to vary within the narrow interval of 0.0218--0.0258\,mag. The average PS is
    described by $\gamma=1.07(5)$ and
    a candidate QPO, resolving into two components, was also detected near
    30\,min.
    
    That this periodicity has not been detected
    so far could be attributed to the following
    reasons. First of all, the photometric studies were performed either
    during the return to normal brightness or during the fading. The state of the CV is unknown only in the study
    performed by \citet{mis}, consisting of only 
    6\,h of observation. In all
    cases, however, there was high amplitude flickering that could easily be the
    reason for the periodicity's masking. 
    Furthermore, \citet{sch} point out that as the system brightened, 
    approaching its normal
    high state, the amount of flickering declined in magnitude
    units. As noted above, they report a decrease in amplitude from
    0.09 to 0.02\,mag. Therefore, if \object {LQ Peg} is caught in a state other than
    its high one, the periodicity will probably be
    masked by the increased flickering. The clear
    appearance of the signal in our data should be connected to
    \object {LQ Peg} being at its normal high-brightness state.
    The steadiness of the flickering amplitude throughout the three-month
    observing time-interval could also be an indication that
    the system is neither fading nor brightening, but lies at its normal
    high state instead. 
    
    Concerning the interpretation of the photometric signal and due to
    its similarity with that of V1193 Ori it could also, for the
    same reasons, represent
    the $P_{\rm orb}$ resulting from the irradiation of the secondary
    star.
    However, if we
    consider that (i) this modulation has come up during the
    system's high-state, (ii) no periodicity had been found in the
    previous years when it was observed in states other than
    its high one, (iii) it has a rather large amplitude with a
    triangular-like form, and (iv) there
    is a suspected spectroscopic $P_{\rm orb}$
    of 2.9\,h, then the photometric signal could be a $P^+_{\rm sh}$.
    However, as previously discussed in Sect. 3.2, the $P_{\rm orb}$ of 
    \object{LQ Peg} is not very accurately known. 
    Even if the 2.9\,h period was correct and the 2.99\,h  was the
    $P^+_{\rm sh}$, the fractional period excess $\epsilon=(P_{\rm sh}-P_{\rm orb})
    / P_{\rm orb}$ would be about 0.03, much
    too small for a system with such a $P_{\rm orb}$
    \citep{pat5}. But given the uncertainty in $P_{\rm orb}$, there is
    nothing to rule out the possibility that 
    the modulation could also be a $P^-_{\rm sh}$, slightly shorter than the $P_{\rm orb}$.
    
    The WHT spectra, both in the blue and the red arms,
    show single-peaked emission lines. All lines are weak compared to the continuum and show no
    orbital radial-velocity variations. 
    Although the spectroscopic data are not adequate, some possible
    explanations 
    could be a disc wind, a
    very low inclination, and emission-line components produced by the
    irradiating side of the secondary. The WHT spectra were obtained during
    a high state of LQ Peg as is deduced from the long-term light
    curve of \citet{hon2}, and this is the state that mostly favours
    the existence of an AD driven wind, due to the high
    mass transfer rate. Such a situation could also have explained 
    how the full widths of the emission
    lines in the spectra we analysed proved to be half those of
    Ringwald and that the lines are now single-peaked. If the wind has grown stronger,
    obscuring the inner radii of the disc from where it originates,
    this could result in a velocity decrease in the wings of our
    lines, giving narrower lines, as well as filling in the space between the
    two disc peaks, giving single-peaked lines.
    In the optical, such winds are revealed via either 
    P-Cygni profile, a shallow, blue-shifted, very broad absorption
    component in the emission lines, or by a red-emission wing of
    a line, characteristic of a receding outflow 
    \citep[for examples see][]{kaf}. However, no such signature was ever detected in our
    set of spectra. Concerning the second explanation, it is true
    that, if LQ Peg is a very low inclined
    system, then stationary single-peaked emission lines would be
    expected. 
    Such a scenario, though, would be in accordance with the powerful
    photometric modulation, only if this signal is a superhump and not
    if caused by irradiation. That is, because irradiation-caused modulations are not expected to be
    seen in low-inclined
    systems, while superhumps are known to be nearly
    independent of inclination \citep{warcv}. 
    The third possible
    explanation has its origin in the chromospheric Balmer emission
    that is commonly observed in CVs \citep{warcv}. The irradiating part of the
    secondary facing the primary has in many cases appeared as a
    narrow emission-line component. That this component creates radial
    velocities approximately in anti-phase to those of the emission lines
    originating from the AD means that the radial
    velocities of the two Balmer lines could include unresolved emission components
    from both the AD and the irradiating part of the
    secondary. But being in anti-phase will result in the computation
    of almost no radial velocities. 

    We believe that in our spectra the
    system lies in a higher state than that of Ringwald, so
    the anti-phased secondary component is enhanced, producing
    negligible radial-velocity variations. However, another possible mechanism
    producing stationary emission has been detected during DN
    outbursts \citep{ste} and is referred to as ``slingshot
    prominences''. During this phenomenon, plasma flows along the
    lines of stable magnetic loops formed by the
    secondary. It then piles up on top of the loops and, confined there, it
    forms a dense region of gas, illuminated by the disc and the
    irradiating face of the secondary. This leads to an emission
    source co-rotating with the secondary, which could
    produce stationary emission-lines. However, in our case this can only remain a
    hypothesis, due to the lack of adequate data.
    %%%%%%%%%%%%%%%%%%%%%%%%%%%%%%%%%%%%%%%%%%%%%%%%%%%%%%%%%%%%%%%%%%%%%%%%%%%%%
    %%%%%%%%%%%%%%%%%%%%%%%%%%%%%%%%%%%%%%%%%%%%%%%%%%%%%%%%%%%%%%%%%%%%%%%%%%%%%
    %%%% \object {LD 317} %%%%%%
    \subsection{\object {LD 317}}
    In 2003, our observations indicate an apparently fading episode,
    which was followed by an
    increase in the flickering amplitude and $\gamma$, in contrast to \object{V1193 Ori}. Therefore, no
    conclusion on the correlation between a system's mean brightness
    and its flickering activity can be drawn from this study.
    No periodic signal was detected from our data, but this might have been caused
    by the very large flickering activity noted. 
    
    Combining the facts that \object {LD 317} has already been proposed as a NL CV,
    that in our observations it has appeared to fade by 1\,mag in an
    interval of approximately 50\,d, and that it appears in an even lower state
    in 2005, we conclude that it should belong to the VY Scl subtype. In
    this respect we must have caught the system during two fading
    episodes, one in 2003 and one in 2005. Judging by its mean magnitudes,
    the star fades by at least 2.5\,mag. 
     
    One NL CV that presents some similarities to \object {LD 317} is
    found to be V794 Aql of the VY Scl subtype. They share an almost
    identical $P_{\rm orb}$, while the slow decline is also in
    common. V794 Aql has been reported to show declines by about 1--3\,mag
    in intervals of 50--100\,d \citep{hon3}.
    Multiple cycles of slow declines
    are  uncommon among NL CVs but a few cases certainly
    exist. Unfortunately the coverage of our runs was insufficient for
    revealing any rising episodes or any data belonging to the system's normal high
    state. In this respect, we are not yet able to conclude much about the
    system's behaviour as a VY Scl star. The only thing,
    though, that can be stated for certain is that the
    system's fadings are associated with declines of up to 2.5\,mag. 
    Moreover, the decline during September--October 2003 is
    considered slow compared to those of
    other VY Scl stars and has a duration of at least 58\,d according to our
    data. This does not, however, exclude shorter durations. As was the case
    with V794 Aql, the fadings could vary both in duration and depth.
    
    Light curves from AAVSO observers before our 2003 data and during Autumn 2004
    show that in October 2003 the system was
    already in a fading episode, but still brighter than when we observed it. 
    It also seems that it started a fading episode in October 2004. At
    that time, it was still much brighter than when we observed it in January 2005.
    
    %%%%%%%%%%%%%%%%%%%%%%%%%%%%%%%%%%%%%%%%%%%%%%%%%%%%%%%%%%%%%%%%%%%%%%%%%%%%%
    %%%%%%%%%%%%%%%%%%%%%%%%%%%%%%%%%%%%%%%%%%%%%%%%%%%%%%%%%%%%%%%%%%%%%%%%%%%%%
    %%%% \object {V795 Her} %%%%%%
     \subsection{\object {V795 Her}}
    %%%%%%%%%%%%%%%%%%%%%%%%%%%%%%%%%%%%%%%%%%%%%%%%%%%%%%%%%%%%%%%%%%%%%%%%%%%%%
    %%%%%%%%%%%%%%%%%%%%%%%%%%%%%%%%%%%%%%%%%%%%%%%%%%%%%%%%%%%%%%%%%%%%%%%%%%%%%
     All resulting light curves
     reveal that the 2.8\,h modulation is present and of high amplitude. We confirm 
     the previously reported QPO near 80\,c\,d$^{-1}$, as well as its resolution into two
     components.
      
    According to \citet{pat1} the signal is definitely
    phase stable on time scales smaller than 20\,d, and an extensive observing
    campaign was proposed to test the longer term stability.
    In our campaign, the frequency of the modulation, the amplitude, and the
    phase did not prove stable. The different values we found agreed with
    different previously reported values. This enhances the speculation
    that the similar but different values reported for the modulation so
    far were expected. This is nevertheless inevitable since this
    modulation is by now strongly and broadly believed to be a superhump
    and superhumps are known for such instabilities. 
    Additionally, if it is a
    superhump then, as also noted by \citet{pat1}, 
    $\epsilon \approx0.08$ which is in
    accordance to the known
    empirical relation between $\epsilon$ and $P_{\rm orb}$ as shown in Fig. 17 of \citet{pat4}.
    We therefore favour the
    disc-precessing model with the $P_{\rm sh}$ being unstable not only in
    period and amplitude but also in phase for time intervals
    longer than 20\,d.

    %%%%%%%%%%%%%%%%%%%%%%%%%%%%%%%%%%%%%%%%%%%%%%%%%%%%%%%%%%%%%%%%%%%%%%%%%%%%%
    %%%%%%%%%%%%%%%%%%%%%%%%%%%%%%%%%%%%%%%%%%%%%%%%%%%%%%%%%%%%%%%%%%%%%%%%%%%%%
    %%%% MCT %%%%%%% 
    \subsection{\object {MCT 2347-3144}}
     The two runs 
     show a difference in the mean magnitude of the system, which appeared brighter in 2003. 
     In 2002, the most likely period 
     is around  6\,h. In 2003, no periodicity could be found. This could, however, 
     be attributed to the increase - by a factor 0.5- in the flickering. Possible QPOs were
     detected near 25 and 50\,c\,d$^{-1}$.
        %%%%%%%%%%%%%%%%%%%%%%%%%%%%%%%%%%%%%%%%%%%%%%%%%%%%%%%%%%%%%%%%%%%%%%%%%%%%%

	\begin{acknowledgements}
	We are grateful for the generous allocations of time to
	Prof. Klaus Reif and Prof. Wilhelm Seggewiss at the Observatorium Hoher List, to Dr. Panayotis
	Boumis at Kryoneri Observatory, to Prof. John Papamastorakis,
	Dr. Iosif Papadakis, and
	Dr. Pablo Reig at Skinakas Observatory. 
	Skinakas Observatory is a collaborative project of the University of
	Crete, the Foundation for Research and Technology-Hellas, and the
	Max--Planck--Institut f\"ur extraterrestrische Physik.
	This paper uses 
	observations made at the South African Astronomical Observatory (SAAO)
	and is partially based on data from the ING Archive. 
	We thank Dr. F.A. Ringwald for kindly providing us with some
	results of his work. We are also grateful to the referee for
	his careful, exhaustive, and very useful report.
	Moreover, this work has been partly supported by ``IAP P5/36'' Interuniversity 
	Attraction Poles Programme of the Belgian Federal Office for 
	Scientific, Technical, and Cultural Affairs.
	C.P. gratefully acknowledges a doctoral
	research  grant by the Belgian Federal Science Policy Office
	(Belspo). J.C. acknowledges financial support from the
	Fund for Scientific Research - Flanders (Belgium).
	\end{acknowledgements}
    \bibliographystyle{aa}
    \bibliography{references}
    
    %\appendix
    \Online
    \appendix
    \section{}
   %%%%%%%%%%%%%%%%%%%%%%%%%%%%%%%%%%%%%%%%%%%%%%%%%%%%%%%%%%%%%%%%%%%%%%%%%%%%%
    \begin {table*}[!ht]
      \caption {Log of observations.}
      \label{t_log_all}
      \centering
      \begin{tabular}{llllll|llllll}
	\hline\hline
	CV & UT date & Site & $HJD_{\rm start}$ & Dur. & Band &
	CV & UT date & Site & $HJD_{\rm start}$ & Dur.& Band\\
	\hline
        V1193 Ori &22oct02 & SAAO & 2570.542 & 1.8  & -   &V795 Her	& 19jul03 & SK   & 2840.291  & 3.0  & $V$ \\  %& 176  & 110
	&23oct02 & SAAO & 2571.504 & 2.7  & -   &---cont.---  & 19jul03 & SK   & 2840.419  & 2.8  & $R$ \\  %& 214  & 100
	&24oct02 & SAAO & 2572.420 & 4.6  & -   &		& 20jul03 & SK   & 2841.274  & 3.0  & $V$ \\  %& 489  & 105   	 
	&25oct02 & SAAO & 2573.436 & 4.2  & -   &		& 21jul03 & SK   & 2842.269  & 3.0  & $V$ \\  %& 470  & 110   		 
	&26oct02 & SAAO & 2574.448 & 3.9  & -   &		& 21jul03 & SK   & 2842.393  & 3.0  & $R$ \\  %& 463  & 110   		 
	&04nov03 & OHL  & 2948.499 & 4.2  & -   &		& 22jul03 & SK   & 2843.265  & 3.0  & $V$ \\  %& 329  & 113   	 
	&05nov03 & OHL  & 2949.472 & 4.9  & -   &		& 22jul03 & SK   & 2843.393  & 3.0  & $R$ \\  %& 221  & 111   		 
	&06nov03 & OHL  & 2950.474 & 4.9  & -   &		& 23jul03 & SK   & 2844.268  & 1.5  & $V$ \\  %& 297  & 120   		 
	&08nov03 & OHL  & 2952.465 & 4.9  & -   &		& 23jul03 & SK   & 2844.405  & 1.5  & $R$ \\  %& 292  & 058   		 
	&17jan04 & OHL  & 3022.333 & 2.9  & -   &		& 11sep03 & OHL  & 2894.316  & 2.8  & -   \\  %& 146  & 282   		 
	&18jan04 & OHL  & 3023.293 & 1.0  & -   &		& 13sep03 & OHL  & 2896.291  & 2.5  & -   \\  %& 044  & 365   		 
	&10oct04 & OHL  & 3289.607 & 2.2  & -   &		& 14may04 & OHL  & 3140.381  & 1.4  & -   \\  %& 129  & 080   	 
	&12oct04 & OHL  & 3291.574 & 2.0  & -   &		& 15may04 & OHL  & 3141.365  & 1.8  & -   \\  %& 166  & 112   	 
	&10jan05 & OHL  & 3381.348 & 1.8  & -   &		& 16may04 & OHL  & 3142.362  & 1.8  & -   \\  %& 084  & 120    
	&14jan05 & OHL  & 3385.382 & 2.8  & -   &		& 19may04 & KR   & 3145.306  & 7.0  & -   \\  %& 250  & 413   	      
	LQ Peg    &01jun04 & SK   & 3158.469 & 2.8  & $R$ &		& 20may04 & KR   & 3146.303  & 7.0  & -   \\  %& 102  & 420 
	&02jun04 & SK   & 3159.465 & 3.2  & $R$ &		& 21may04 & KR   & 3147.312  & 6.4  & -   \\  %& 112  & 617 
	&03jun04 & SK   & 3160.466 & 3.2  & $R$ &		& 05jun04 & SK   & 3162.349  & 2.3  & $R$ \\  %& 121  & 091 
	&05jun04 & SK   & 3162.454 & 3.1  & $R$ &		& 14jan05 & OHL  & 3385.673  & 2.1  & -   \\  %& 112  & 185
	&06jun04 & SK   & 3163.455 & 3.6  & $R$ &		& 03may05 & KR   & 3494.369  & 2.9  & -   \\  %& 081  & 312
	&19aug04 & KR   & 3237.363 & 5.0  & -   &		& 04may05 & KR   & 3495.350  & 2.3  & -   \\  %& 316  & 137
	&20aug04 & KR   & 3238.287 & 7.7  & -   &		& 05may05 & KR   & 3496.345  & 5.9  & -   \\  %& 380  & 108
	&21aug04 & KR   & 3239.345 & 6.5  & -   &		& 07may05 & KR   & 3498.336  & 2.5  & -   \\  %& 324  & 095
	&22aug04 & KR   & 3240.454 & 3.5  & -   &		& 08may05 & KR   & 3499.337  & 2.7  & -   \\  %& 225  & 124
	&30aug04 & SK   & 3248.397 & 6.8  & $R$ &		& 09may05 & KR   & 3500.323  & 3.1  & -   \\  %& 245  & 232
        LD 317    &11sep03 & OHL  & 2894.468 & 0.5  & -   &		& 11may05 & KR   & 3502.353  & 2.0  & -   \\  %& 061  & 086
	&13sep03 & OHL  & 2896.419 & 3.8  & -   &		& 13may05 & KR   & 3504.333  & 2.9  & -   \\  %& 453  & 205
	&14sep03 & OHL  & 2897.291 & 6.0  & -   &		& 14may05 & KR   & 3505.441  & 1.3  & -   \\  %& 765  & 080
	&04nov03 & OHL  & 2948.248 & 5.4  & -   &   MCT	& 24oct02 & SAAO & 2572.290  & 3.0  & -   \\  %& 360  & 231
	&05nov03 & OHL  & 2949.232 & 5.0  & -   & 2347-3144	& 25oct02 & SAAO & 2573.305  & 3.0  & -   \\  %& 453  & 253  
	&06nov03 & OHL  & 2950.222 & 6.0  & -   &		& 26oct02 & SAAO & 2574.277  & 3.6  & -   \\  %& 520  & 285
	&08nov03 & OHL  & 2952.345 & 2.8  & -   &		& 27oct02 & SAAO & 2575.272  & 4.2  & -   \\  %& 160  & 293
	&12jan05 & OHL  & 3383.240 & 3.0  & -   &		& 13aug03 & SAAO & 2865.398  & 4.5  & -   \\  %& 221  & 083
	V795 Her    &02jun03 & KR   & 2793.477 & 2.5  & -   &		& 16aug03 & SAAO & 2868.343  & 8.5  & -   \\  %& 220  & 143
	&03jun03 & KR   & 2794.342 & 6.0  & -   &		& 17aug03 & SAAO & 2869.379  & 7.0  & -   \\  %& 842  & 148   									     
	\hline									   
      \end{tabular}
	  {\footnotesize 
	    \newline 
	    Notes: $HJD_{\rm start}=HJD-2450000$; Dur. is the duration of
	    each run in hours.\hfill}
    \end {table*}
    %%%%%%%%%%%%%%%%%%%%%%%%%%%%%%%%%%%%%%%%%%%%%%%%%%%%%%%%%%%%%%%%%%%%%%%%%%%%%%
    \begin{figure*}
      \centering
      \includegraphics[width=17cm]{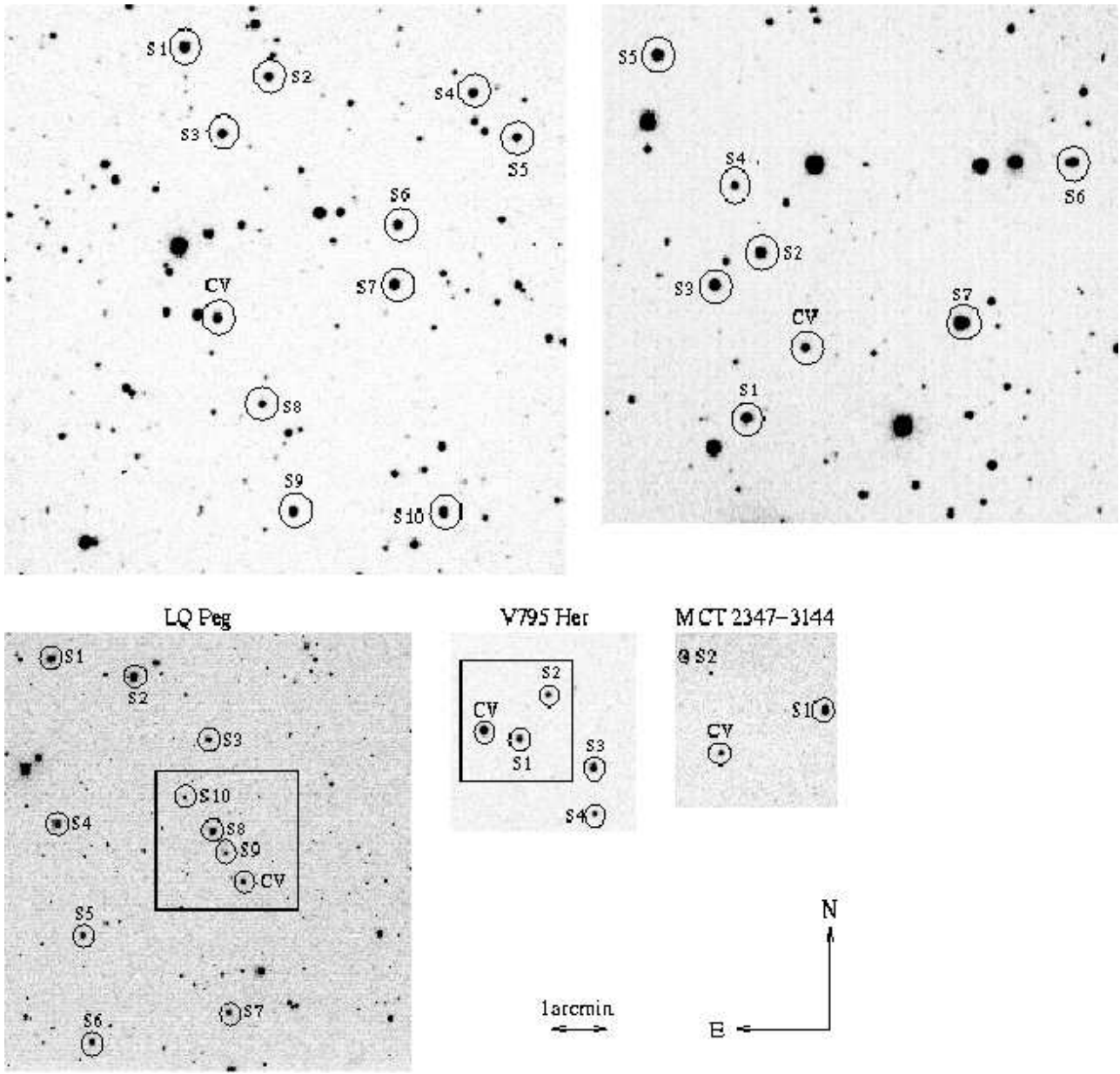}
      \caption{Finding charts of the five selected CVs, along with the
	selected comparisons. When present, boxes correspond to
	observing runs with a smaller FOV.}
      \label{f_fov_all}
    \end{figure*}
    %%%%%%%%%%%%%%%%%%%%%%%%%%%%%%%%%%%%%%%%%%%%%%%%%%%%%%%%%%%%%%%%%%%%%%%%%%%%% 
    \begin{figure*}
      \centering
      \includegraphics[width=17.5cm]{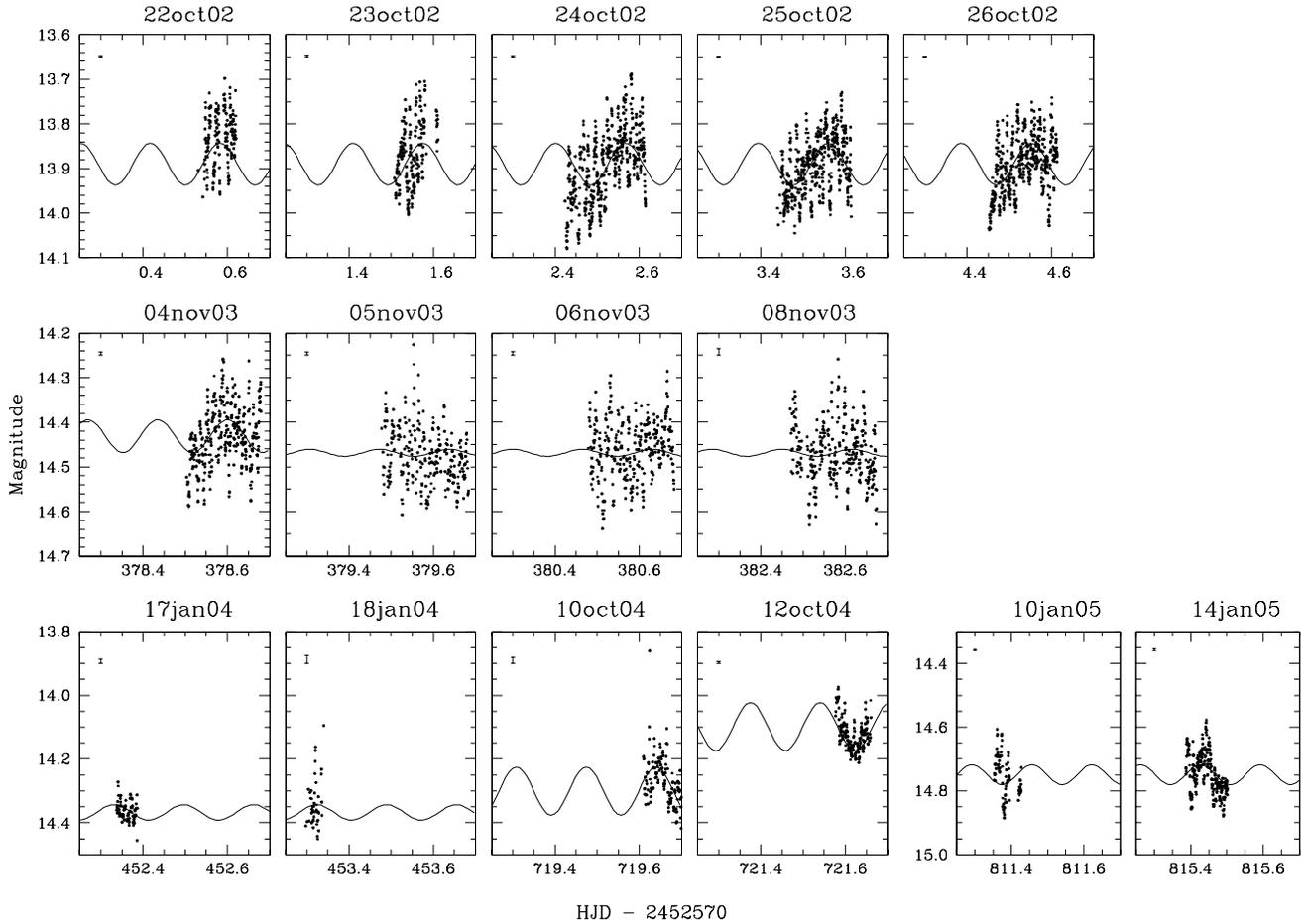}
      \caption{\object {V1193 Ori} light curves. The periodicity of
	3.97\,h,
	detected in the SAAO data, has been superimposed. The error
	bars in the upper left part of each light curve represent
	the mean $\sigma$ of the comparison stars.}
      \label{f_sin_V1193Ori}
    \end{figure*}   
   %%%%%%%%%%%%%%%%%%%%%%%%%%%%%%%%%%%%%%%%%%%%%%%%%%%%%%%%%%%%%%%%%%%%%%%%%%%%%      
    %%%%%%%%%%%%%%%%%%%%%%%%%%%%%%%%%%%%%%%%%%%%%%%%%%%%%%%%%%%%%%%%%%%%%%%%%%%%%
    \begin{figure*}
      \centering
      \includegraphics[width=17cm]{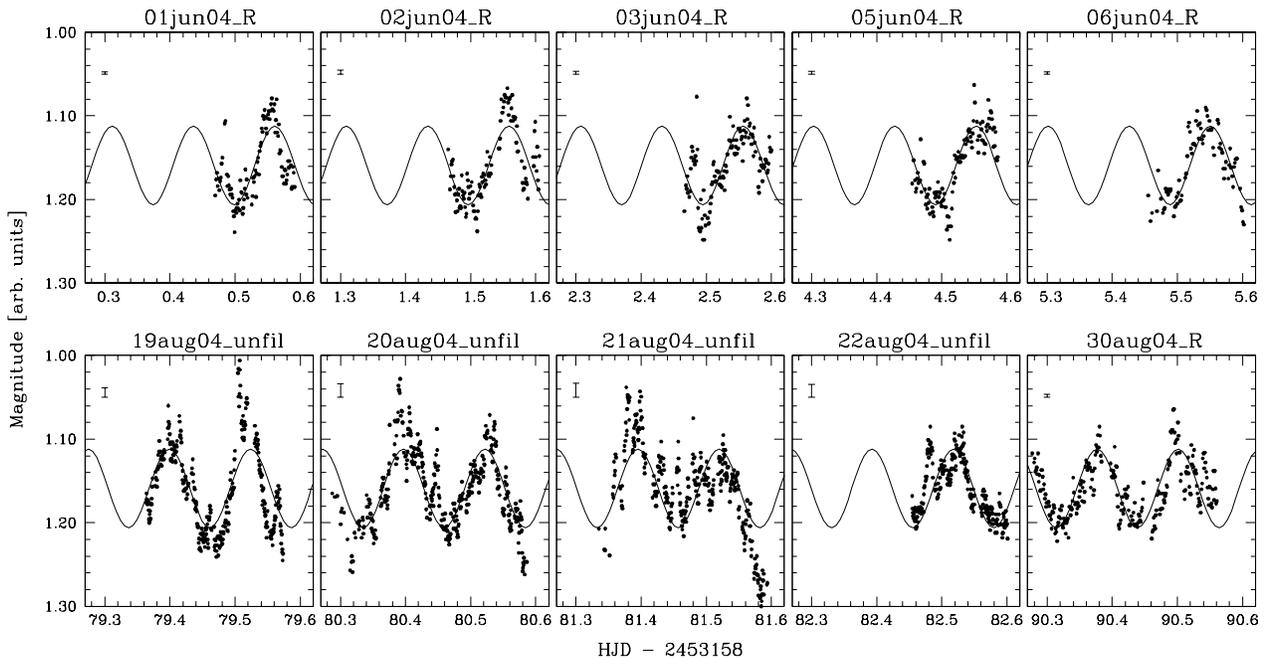}
      \caption{\object {LQ Peg} light curves with the periodicity of
	2.99\,h
	superimposed. The error bars in the upper left part of each
	light curve represent the mean $\sigma$ of the comparison stars.}
      \label{f_sin_LQPeg}
    \end{figure*}
    %%%%%%%%%%%%%%%%%%%%%%%%%%%%%%%%%%%%%%%%%%%%%%%%%%%%%%%%%%%%%%%%%%%%%%%%%%%%%
    %%%%%%%%%%%%%%%%%%%%%%%%%%%%%%%%%%%%%%%%%%%%%%%%%%%%%%%%%%%%%%%%%%%%%%%%%%%%%  
    \begin{figure}
      \resizebox{\hsize}{!}{\includegraphics[width=8cm]{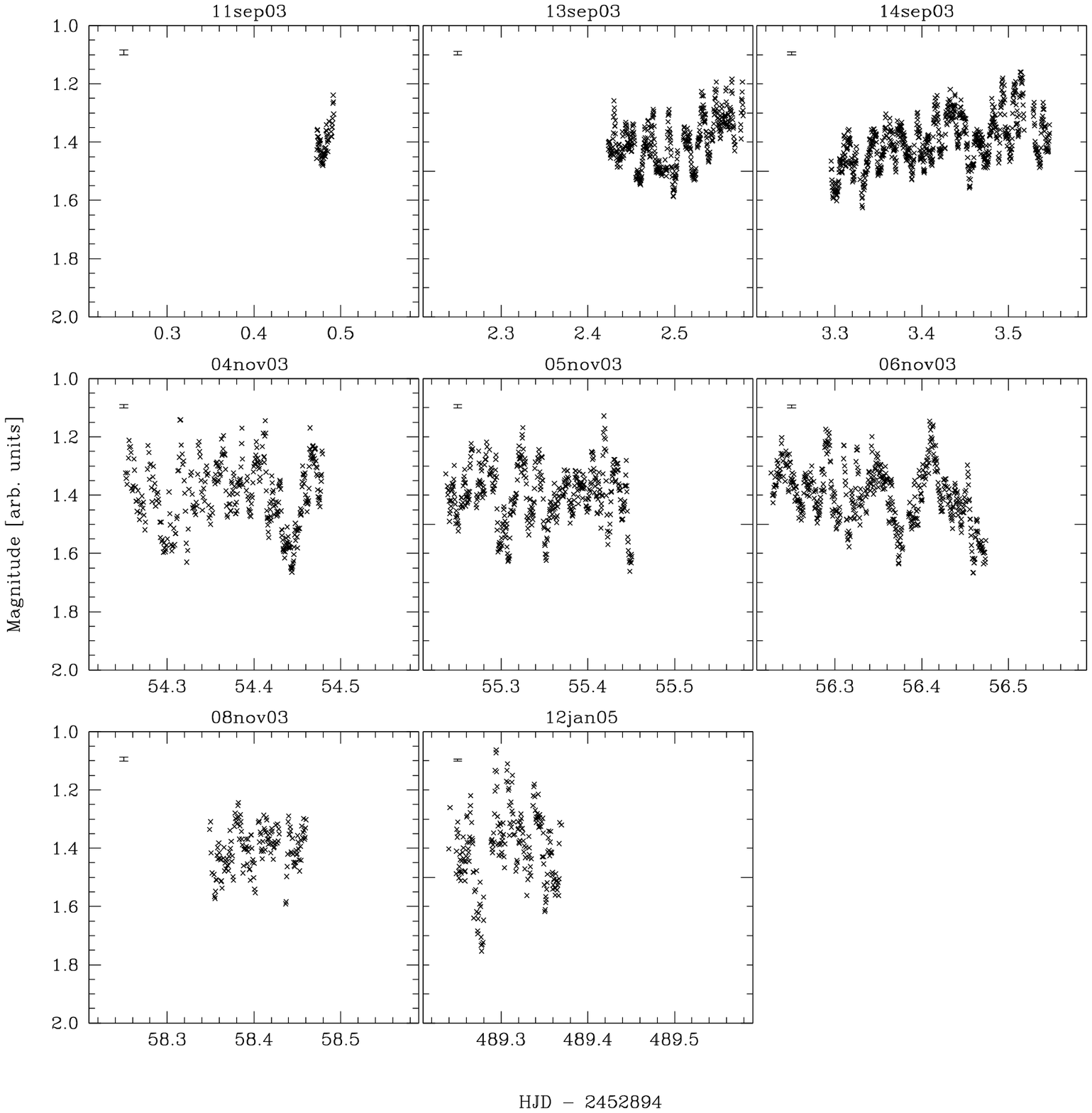}}
      \caption{\object {LD 317} light curves around their nightly mean. The error
	bars in the upper left part of each
	light curve represent the mean $\sigma$ of the comparison stars.}
      \label{f_lc_LD317}
    \end{figure}
    %%%%%%%%%%%%%%%%%%%%%%%%%%%%%%%%%%%%%%%%%%%%%%%%%%%%%%%%%%%%%%%%%%%%%%%%%%%%% 
    %%%%%%%%%%%%%%%%%%%%%%%%%%%%%%%%%%%%%%%%%%%%%%%%%%%%%%%%%%%%%%%%%%%%%%%%%%%%%
    \begin{figure*}
      \centering
      \includegraphics[width=18cm]{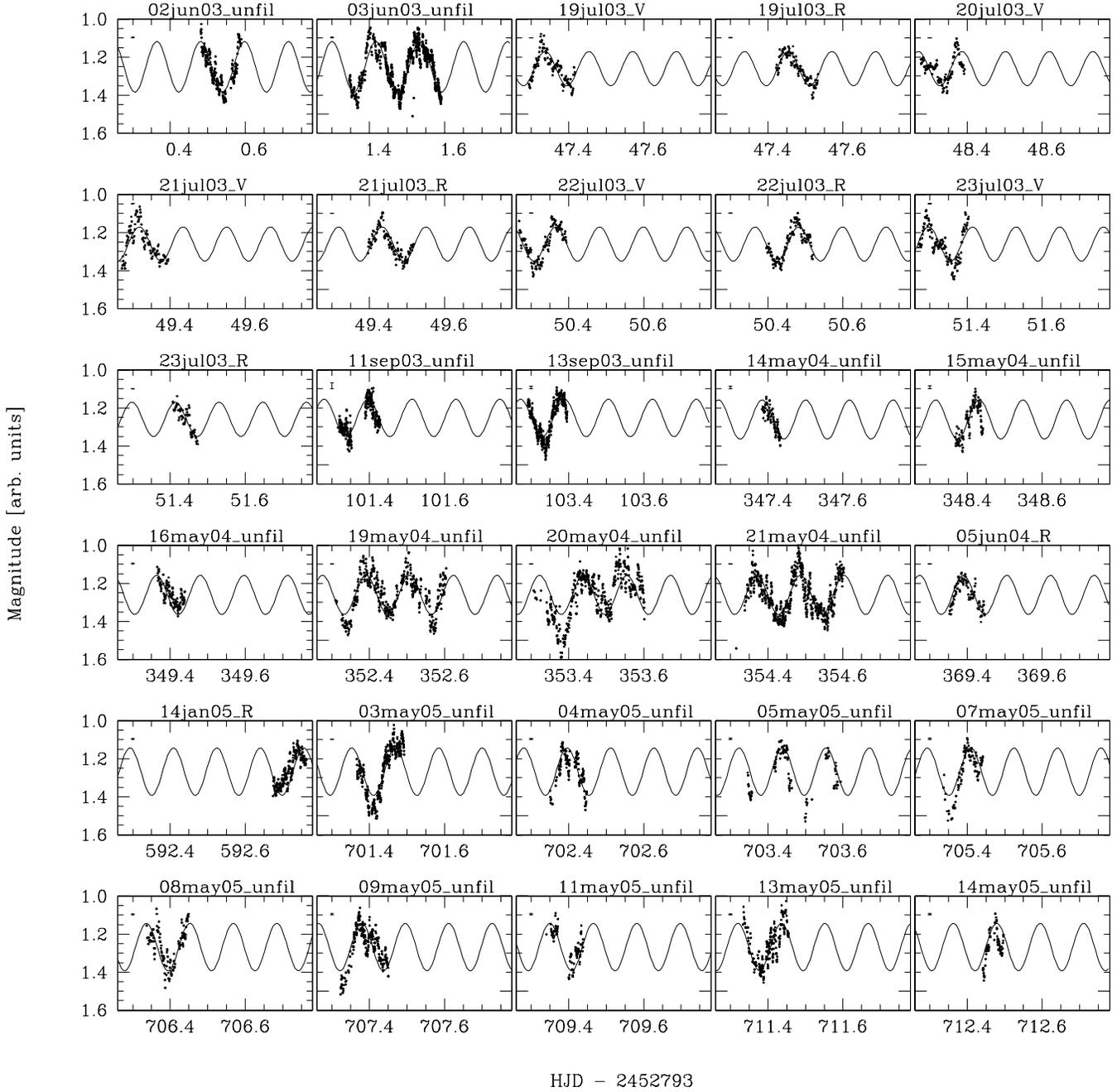}
      \caption{\object {V795 Her} light curves with night-to-night variations removed.
      Superimposed on the light curves of each subgroup has been the sinusoidal 
      representation of the corresponding subgroup (see Table~\ref{t_sub_V795Her}). The error bars in the upper left part of each
	light curve represent the mean $\sigma$ of the comparison stars.}
      \label{f_sin_V795Her}
    \end{figure*}  
    %%%%%%%%%%%%%%%%%%%%%%%%%%%%%%%%%%%%%%%%%%%%%%%%%%%%%%%%%%%%%%%%%%%%%%%%%%%%%
    %%%%%%%%%%%%%%%%%%%%%%%%%%%%%%%%%%%%%%%%%%%%%%%%%%%%%%%%%%%%%%%%%%%%%%%%%%%%%
    \begin{figure}
      \resizebox{\hsize}{!}{\includegraphics[width=8cm]{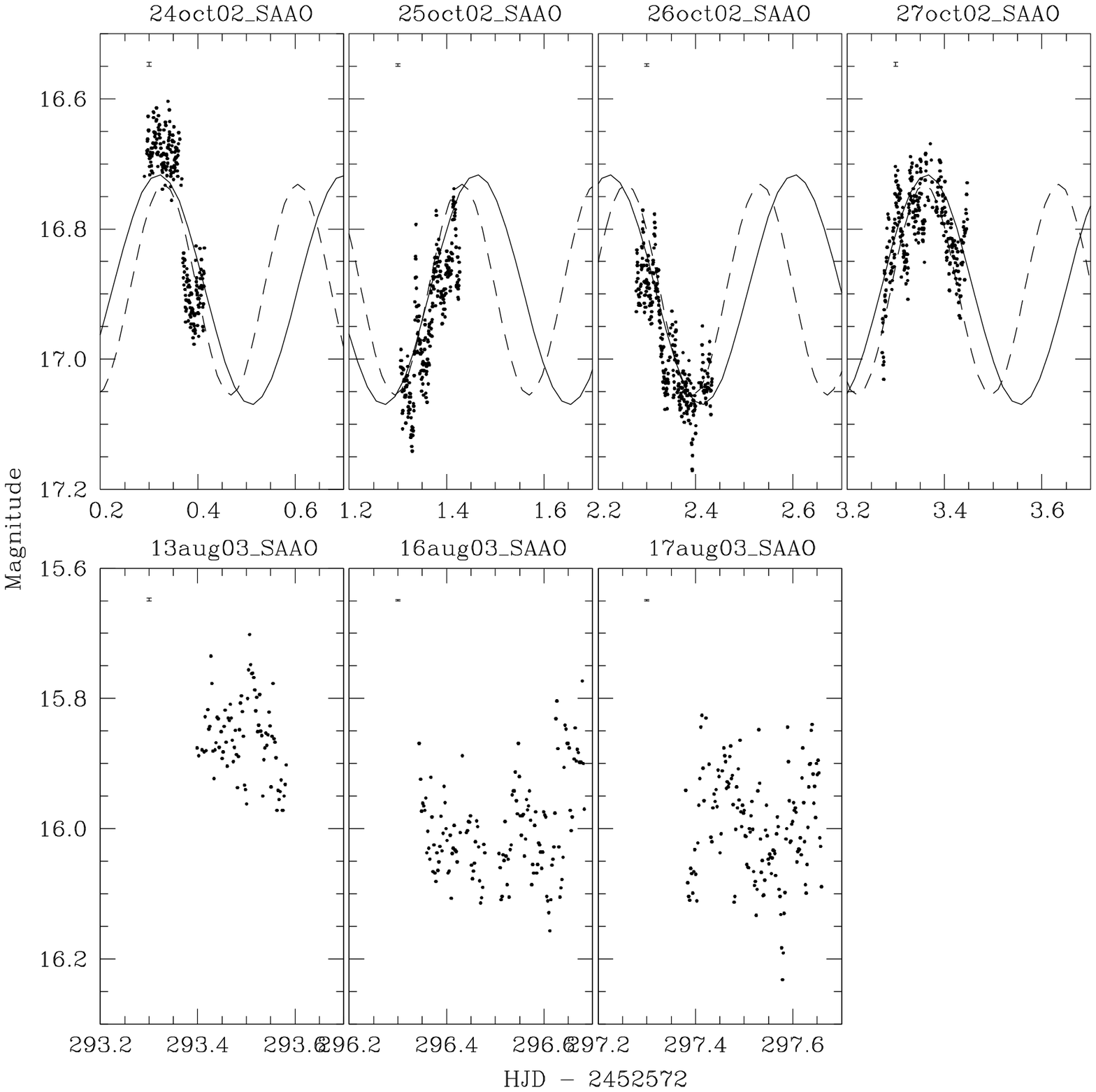}}
      \caption{\object {MCT 2347-3144} light curves. Superimposed on the 2002
	data are the two most probable periodicities. The full line
	corresponds to 2.6\,c\,d$^{-1}$, while the dashed line to 3.6\,c\,d$^{-1}$.
	The error bars in the upper left part of each
	light curve represent the mean $\sigma$ of the comparison stars.}
      \label{f_sin_Scl}
    \end{figure}
    %%%%%%%%%%%%%%%%%%%%%%%%%%%%%%%%%%%%%%%%%%%%%%%%%%%%%%%%%%%%%%%%%%%%%%%%%%%%%
\end{document}